\documentclass[lettersize, journal]{IEEEtran}
\usepackage{amsmath, graphicx}
\usepackage{graphicx}
\usepackage{algorithm}
\usepackage[utf8]{inputenc}
\usepackage[T1]{fontenc}
\usepackage{boldline}

\usepackage{amsmath}
\usepackage{amssymb}

\usepackage{xcolor} 

\usepackage{hyperref}
\usepackage{enumitem}
\usepackage{dsfont} 
\usepackage{wrapfig}

\setlist{leftmargin=3mm}

\usepackage{tabularx}
\usepackage{multirow}
\usepackage{ifthen}
\usepackage{xargs}     
\usepackage{footnote}
\makesavenoteenv{tabular}
\usepackage{threeparttable}

\newcommand\subfigwidth{0.19}
\newcommand\subfigwidthfour{0.24}

\usepackage[disable]{todonotes} 

\newcommand{\bd}[1]{\textbf{#1}}

\usepackage{cleveref}

\usepackage[caption=false,font=footnotesize,subrefformat=parens]{subfig}
\usepackage{xspace}

\def\ie{{\textit{i.e.}\xspace}} \def\eg{{\textit{e.g.}\xspace}}
\def\aka{{\textit{a.k.a.}\xspace}} 
\def\sysname{{\sc RadioSES}\xspace}
\def\dlmodule{{\sc RadioSESNet}\xspace}
 
 \newcommand{\head}[1]{{\noindent \bf #1}}

\newcommand{\ar}{AR}
\newcommand*{\myalign}[2]{\multicolumn{1}{#1}{#2}}

\def\metricone{{\footnotesize SiSDR}} 
\def\metrictwo{{\footnotesize SIR}} 
\def\metricthree{{\footnotesize STOI}}
\def\metricfour{{\footnotesize PESQ}}
\def\S{Section~}

\begin{document}

\title{\sysname: mmWave-Based Audioradio Speech Enhancement and Separation System}

\author{Muhammed~Zahid~Ozturk,~\IEEEmembership{Student~Member,~IEEE,}
Chenshu~Wu,~\IEEEmembership{Senior~Member,~IEEE,}
Beibei~Wang, ~\IEEEmembership{Senior~Member,~IEEE,}
Min~Wu,~\IEEEmembership{Fellow,~IEEE,}
and~K.~J.~Ray~Liu,~\IEEEmembership{Fellow,~IEEE}

\thanks{M. Z. Ozturk, B. Wang, and K. J. R. Liu are with the Department
of Electrical and Computer Engineering, University of Maryland, College Park,
College Park, MD 20742 USA, and also with Origin Wireless Inc., Greenbelt,
MD 20770, USA. C. Wu is with the Department of Computer Science, University of
Hong Kong, and also with Origin Wireless Inc. (e-mail: ozturk@umd.edu, chenshu@cs.hku.hk, bebewang@umd.edu,
minwu@umd.edu, kjrliu@umd.edu).}
}

\maketitle

\begin{abstract}
Speech enhancement and separation have been a long-standing problem, especially
with the 
recent
advances using a single microphone. Although microphones perform well in
constrained settings, their performance for speech
separation decreases in noisy conditions. In this work, we propose \sysname, an
audioradio speech enhancement and separation system that overcomes inherent problems in audio-only systems. 
By fusing a complementary radio modality, \sysname can estimate the number of speakers, solve source
association problem, separate and enhance noisy mixture
speeches, and improve both intelligibility and perceptual quality. 
We perform millimeter-wave sensing to detect and localize speakers, 
and introduce an audioradio deep learning framework to fuse the separate radio features with the mixed audio features. 
Extensive experiments using commercial off-the-shelf devices show that \sysname outperforms a
variety of state-of-the-art baselines, with consistent performance
gains in different environmental settings. 
Compared with
the audiovisual methods, \sysname provides similar improvements
(\eg~3 dB gains in SiSDR), 
along with the benefits of lower computational complexity and being less privacy
concerning. 

\end{abstract}

\section{Introduction}
\label[]{sec:intro}
Humans are enormously capable of understanding a noisy speech 
or separating one speaker from
another, we collectively refer to these capabilities as speech enhancement and separation (SES), and is
known as the cocktail party problem \cite{cherry1953some}.
SES capability for computers is of great demand for many applications, such as voice commands, 
live speech recording, etc., yet remains a challenging problem using microphones. 

Monaural SES methods achieved remarkable progress in the recent years
with the help of deep learning, especially when there is not much background
noise \cite{wichern2019wham}. 
However, fundamental problems still exist in estimating the number of sources in a mixture, associating
output sources with the desired speakers (\aka~label permutation problem), and tracing the speakers for long periods of time. 
Although
these problems can be solved for clean mixtures, by clustering-based methods \cite{hershey2016deep} and
permutation invariant training (PIT) \cite{yu2017permutation}, 
their performance can decrease with noisy mixtures. 
Overall, audio-only approaches suffer from these ill-posed problems inherently. 

To overcome the problems and enhance SES, 
multimodal systems have been introduced to exploit readily available information beyond audio, such as video
\cite{ephrat2018looking,afouras2018conversation}. 
Similar
to human perception, which also uses lip motion and facial information
\cite{golumbic2013visual}, 
audiovisual 
systems are
shown to improve SES performance, especially
in challenging cases, such as same-speaker mixtures. Same and similar-speaker
mixtures are especially difficult for audio-only methods, as the distinction
between the two sources is minimal. 
Additional visual information about the speaker, e.g., videos or even a facial picture
of the user \cite{chung2020facefilter}, or other information, such as 
voice activity detection
\cite{rivet2007visual}, or pitch \cite{zhang2016pairwise} improves the SES performance.
However, camera-based methods require good lighting conditions 
and raise potential privacy concerns.

\begin{figure}
	\centering
	\includegraphics[width=1.0\columnwidth]{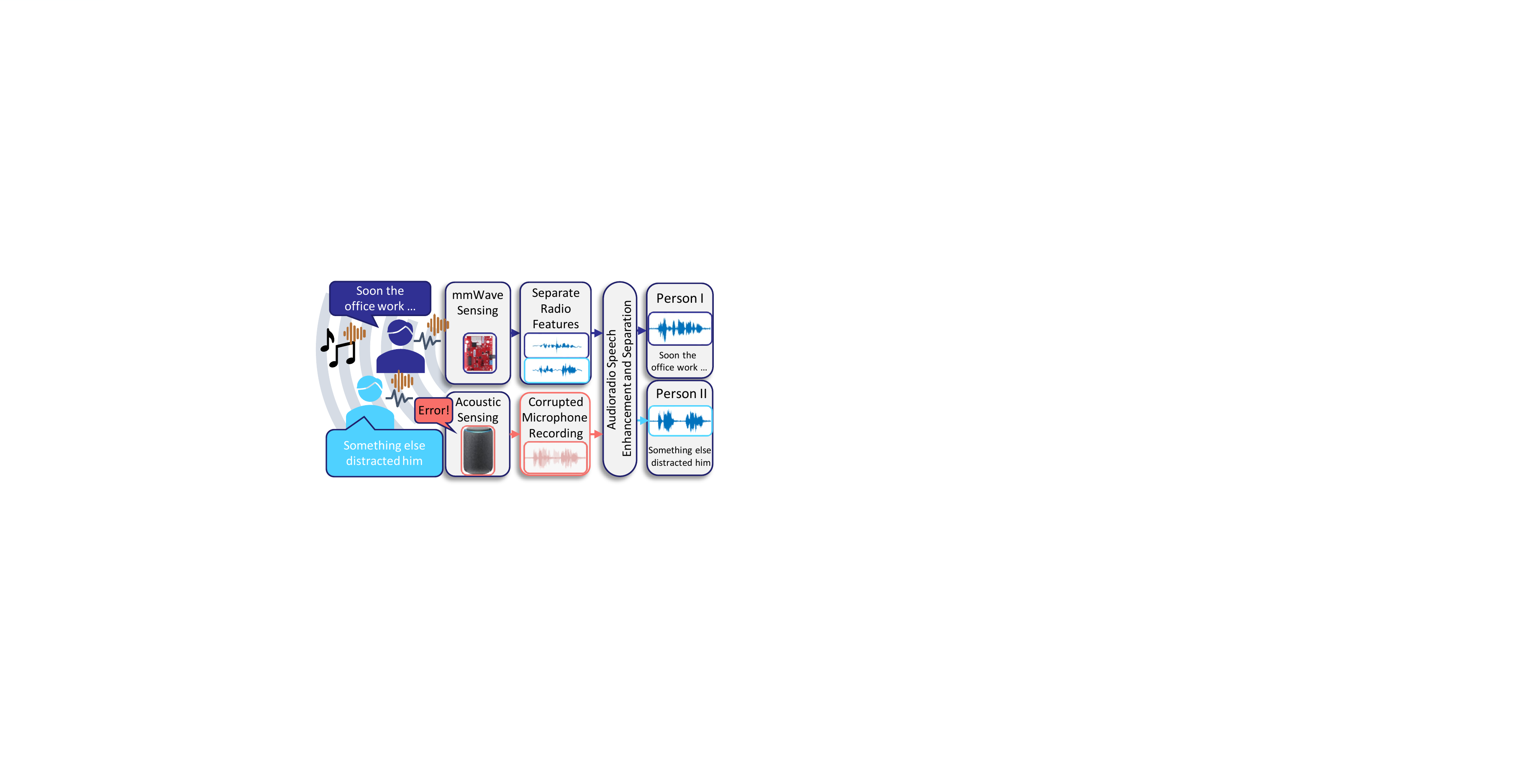}
	\caption{\sysname Overview}
	\label{fig:systemoverview}
\end{figure}

In this work, we propose to address the SES problem by jointly 
leveraging millimeter-wave (mmWave) sensing 
as an orthogonal radio modality. 
Compared to cameras, radio devices are low-power, can operate in dark, through-wall
settings and are less privacy-invasive. 
The radio reflections from speakers not only can allow separation of multiple speakers 
but also capture articulatory motions for SES. 
The reasons to select mmWave radios are two-fold: 
On the one hand, more and more smart devices now include an mmWave radar and a microphone, such as
Google Soli phone and Nest Hub \cite{soli2021sleep,soliphone}, Amazon Alexa \cite{alexa2021radar} etc. 
mmWave sensing promises to be more ubiquitous in
the future. 
On the other hand, 
mmWave sensing has enabled
many applications related to motion and vibration, such as heart rate
monitoring \cite{wang2021vimo}, measuring machinery and object vibration
\cite{ozturk2021sound,jiang2020mmvib}, or extracting vocal folds vibration
\cite{chen2017detection}. 
In particular, it
has been used to estimate pitch and detect voice
activity \cite{chen2017detection}, reconstruct speech to some extent
\cite{xu2019waveear,ozturk2021radiomic}, 
as well as enhance speech recognition for a single speaker \cite{liu2021wavoice}. 
Yet no existing work has explored utilizing both modalities for \textit{joint} SES tasks. 

With this motivation, we develop an \textit{audioradio}\footnote[1]{We combine
audio and radio words as \textit{audioradio} to refer to a multimodal system
consisting of both modalities, similar to the word \textit{audiovisual}.} speech enhancement and
separation system to solve the aforementioned problems and improve the
overall
performance. 
Building an audioradio SES system faces multiple challenges. First,
in order to solve the number of sources problem, 
a robust and efficient
source detection and tracking method is needed, as the performance of a system
can decrease significantly in the event of miss detection. Second, radio
signals are usually prone to environmental effects, and their performance can
decrease considerably when tested at a new location. Returned signals
from the objects are not
only affected by vibration, but also from motion, with motion usually being the
stronger effect.
Third, different from the
rich literature in audiovisual deep learning methods, radio modality has not
been explored in the context of SES. 
Designing a suitable and efficient deep learning model for practical
applications is non-trivial. Last, deep learning systems require extensive data collection and
robust training methods, which is especially challenging for radio signals.

We overcome these challenges in \sysname, the first Audio-\textbf{Radio} \textbf{S}peech \textbf{E}nhancement and
\textbf{S}eparation system. 
As illustrated in Fig. \ref{fig:systemoverview}, \sysname can detect, localize, and estimate the number of
sources in an environment and improve SES performance even in unseen/challenging 
conditions.
To achieve robust detection and localization, we first develop a computationally efficient pipeline 
of signal processing that can extract the radio features for speakers separately. 
Then we design an audioradio deep learning framework that takes both audio and radio signals
as the inputs and outputs separated and enhanced speeches for each of the speakers. 
Following recent advances in monaural SES, our deep learning module, called \dlmodule, 
utilizes adaptive encoders, instead of relying on classical Short-Term Fourier Transform (STFT) representation.
We further introduce a variety of techniques learned from audiovisual SES to improve robustness and generalizability of \dlmodule to 
unseen environments and users. 

We evaluate \sysname using a commercial off-the-shelf (COTS) mmWave radar using 
synthetic and real-world data. 
To boost data collection for training, we build a data
collection platform, and capture 5700 sentences from 19 users.
Our results show that 
the radio modality can complement audio and bring similar improvements
to that of video modality while not imposing visual privacy issues. 
We extensively test \sysname in different number
of mixtures and a variety of environmental
settings.
When compared to the state-of-the art audio-only
method (\eg, DPRNN-TasNet \cite{luo2020dual}), 
\sysname brings around 3 dB improvements for
separating noisy mixtures, along with benefits of estimating the 
number of sources and
associating output streams. The improvements 
are not only in terms of SDR, but also of intelligibility and perceptual quality.
Our results
indicate that audioradio methods have a tremendous potential for SES tasks, as they
enable a low-complexity, effective, privacy-preserving alternative to audio-only or
vision-based methods. \sysname explores an important step in this direction and
will inspire follow-up research. Some experimental results of \sysname are available on our project website: {\url{https://zahidozt.github.io/RadioSES/}

In addition to our preliminary work \cite{ozturk2022toward} that explores the
feasibility of
\textit{audioradio} speech enhancement and separation, our main contributions in
this work are:
\begin{itemize}[leftmargin=*]
	\item We propose \sysname, a novel end-to-end audioradio system that jointly leverages mmWave radio and audio signals
	for simultaneous speech enhancement and separation. 
	\item We introduce an audioradio deep learning framework that fuses audio signals 
	and radio signals for multi-modal speech separation and enhancement. 
    \item We utilize adaptive encoders for time-frequency representation, perhaps for the first time, not only for audio, but also for radio signals without relying on the commonly used spectrograms.
	\item We build an extensive audioradio dataset 
	and compare \sysname's performance in various conditions with
	state-of-the-art methods. \sysname achieves 3 to 6 dB SiSDR improvements in
	separating two and three person mixtures, respectively.
\end{itemize}

The rest of the paper follows a literature review in \S\ref{sec:relatedwork},
and a preliminary in \S\ref{sec:preliminary}.
\S\ref{sec:systemoverview} presents an overview, with detailed design in \S\ref{sec:radio-preprocessing} and
\S\ref{sec:deeplearning}.
We give dataset and implementation
details in \S\ref{sec:implementation}, and present the results in
\S\ref{sec:evaluation}. Last, we discuss in
\S\ref{sec:discussion} and conclude in \S\ref{sec:conclusion}.

\section{Related Work}
\label[]{sec:relatedwork}
 
\head{Audio-only Methods}
\label[]{subsec:related-audioonly}
Traditional methods, such as computational
auditory scene analysis (CASA) \cite{brown1994computational} with
pitch-estimation based separation \cite{hu2010tandem}, nonnegative matrix factorization (NMF) \cite{schmidt2006single}, or
probabilistic methods \cite{virtanen2006speech} cannot
generalize well to unseen speakers \cite{wang2018supervised}, which is a major
limiting factor for their performance. 

Deep learning based methods outperformed classical approaches recently 
\cite{wang2018supervised}. Instead of estimating output representation directly,
these
methods usually estimate a mask that is multiplied with the input. 
Some masks used as training targets are binary mask \cite{hu2001speech}, 
STFT spectral mask \cite{wang2014training}, and complex ratio mask
\cite{williamson2016complex}. PHASEN \cite{yin2020phasen} estimates amplitude
and phase masks separately. 
In SEGAN \cite{pascual2017segan}, a time-domain SE system using generative
adversarial networks has been proposed.  
ConvTasNet \cite{luo2019conv} performs better
than ideal 
ratio mask for SS, with an adaptive/learnable encoder, instead of
classical STFT. Later on, 
the fully convolutional layers in \cite{luo2019conv} are replaced by 
dual-path
RNN
(DPRNN-TasNet)
\cite{luo2020dual}, dual-path transformer network (DPTNET) \cite{chen2020dual},
and fully attention layers in SepFormer
\cite{subakan2021attention}.  

Source association and tracking problems can be solved with frame-level PIT
\cite{yu2017permutation} and utterance-level PIT \cite{kolbaek2017multitalker}.
Even though these methods mitigate the problem, and estimate the same speaker's
speech for a given frame, they can fail when the speakers have similar
pitch and speaking characteristics \cite{lu2019audiovisual}. The number of 
sources can be estimated by deep clustering \cite{hershey2016deep} or deep 
attractor
networks \cite{chen2017deep}. 
However, these models still have the
source tracking problem over long time, which is started to be addressed
recently \cite{nachmani2020voice}.

\head{Multimodal Methods}
\label[]{subsec:related-multimodal}
Vision-based works use different
features as input, such as face
embeddings \cite{ephrat2018looking}, lip embeddings
\cite{afouras2018conversation} or optical flow \cite{lu2019audiovisual}. These
methods
use STFT representation, although
time-domain processing is also possible \cite{wu2019time}.
\cite{oh2019speech2face} estimates faces from the speech signals, whereas
\cite{chung2020facefilter} uses picture of a speaker for separation.
Audiovisual methods
have complex processing pipelines, require good lighting, and raise privacy
concerns.

On the other hand, ultrasound can also be used for
speech generation \cite{toth2010synthesizing}, 
speaker recognition \cite{kalgaonkar2008ultrasonic}, and speech enhancement
\cite{lee2019speech}. UltraSE \cite{sun2021ultrase} uses a deep learning to enhance single
speaker signals. Ultrasound
signals can only work at a short range (\eg~15cm), and are too
coarse to measure fine-grained vocal folds vibration \cite{sun2021ultrase}.

\begin{figure}
	\centering
	\includegraphics[width=0.7\columnwidth]{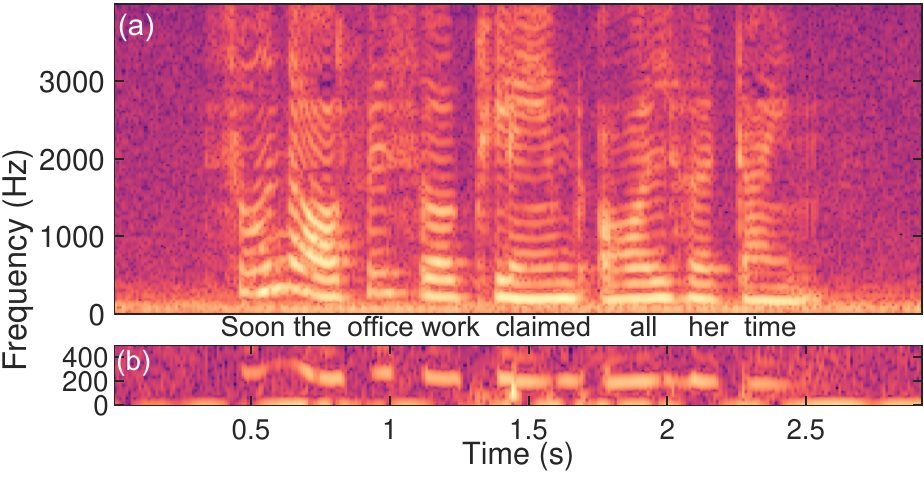}
	\caption{a) Spectrogram of speech, captured with a microphone and sampled at 8 kHz, b) Spectrogram of radio signal, captured from vocal fold's of the speaker in a)}
	\label{fig:audioradio}
\end{figure}

\head{Wireless Sensing}
\label[]{subsec:related-wireless}
Recently, wireless sensing has been an emerging phenomenon
\cite{wang2018promise, liu2019wireless}, with multiple applications to gait monitoring
\cite{yang2020muid,ozturk2021gaitcube}, 
vibration monitoring \cite{jiang2020mmvib} and vital signs
monitoring \cite{wang2021vimo,adib2015smart,yang2016monitoring}. mmWave devices
have enabled
monitoring of $\mu m$-level displacement on object surfaces, and can
capture vocal folds vibration remotely \cite{chen2017detection}.
Furthermore, they can 
recover sound from human throats \cite{xu2019waveear}, speaker diaphragms \cite{uwhear}, or passive
object surfaces, such as a piece of aluminum \cite{ozturk2021radiomic}.
Recently, mmWave 
information from the speaker has also been used for speaker verification
\cite{dong2020secure} or speaker recognition 
\cite{li2020vocalprint}. Although these works can reject
interfering sound by 
sensing its \textit{source}, they capture
\textit{limited} and low-quality sound. 
Such systems do not employ one of the most common sensors available,
microphones, to further improve the quality. A recent work, WaVoice
\cite{liu2021wavoice},
fuses both modalities for speech recognition, but can only recognize commands of
a single user, and is
not suitable for speech separation. To that end, \sysname explores a multimodal
speech enhancement and separation system using radio signals. Furthermore,
to the best of our knowledge, \sysname is the first work to utilize adaptive
encoders for time-frequency representation, instead of spectrograms, which is
the typical method for all radio-based sensing systems.

\section{Preliminary}
\label[]{sec:preliminary}
In this section, we start with an illustration to explain what radio devices
measure. Channel-impulse response (CIR) of a radio device is affected by the 
motion in the environment. Human vocal folds create $\mu m$ level vibration
displacement on the surface of the human body, especially in the throat region,
and 
this displacement changes the amplitude and phase of the returned complex-valued
radar
signals. 
As shown in Fig. \ref{fig:audioradio}, the low-frequency component of the radio
captured spectrogram and microphone captured spectrogram are extremely similar,
as the two modalities measure the same \textit{mechanical vibration}. Radio
devices potentially enable measuring voice activity (as the silence instants do
not include vibration), and pitch tracking. As it will be shown later, this
information from radio signals will be combined with the \textit{corrupted}
audio signals for high quality speech enhancement and separation. We note that,
although Fig. \ref{fig:audioradio} includes spectrograms for illustration,
\sysname uses learnable encoders for time-frequency representation of both audio
and radio modalities.

\section{System Overview}
\label[]{sec:systemoverview}
As an overview, \sysname requires a device with mmWave sensing capabilities, and
a microphone, (\eg \cite{soliphone,soli2021sleep}). 
The monaural microphone records ambient sound,
and
the mmWave radar
is expected to output separate streams for each sound source, where we constrain
our
investigation to speech signals. Although it is possible to place radar in a
separate location, we assume the radar and microphone to be colocated, as in
\cite{soli2021sleep}. We expect the speaking objects to be in front of the
radar. 
In
addition, although radars can sense in NLOS conditions, we only investigate LOS in this work
as our goal is not to eavesdrop. The application scenarios of \sysname can be one or more 
persons speaking in front of a computer, smart hub, or a phone, with LOS. 

Having speaking persons in the field-of-view (FoV), \sysname detects near
stationary bodies and uses the output to estimate
and associate sources with the extracted sound signals. 
Unlike microphone arrays, using mmWave sensing
enables to capture individual data
streams not only from different azimuth
angles, but also from varying distances.
After these tasks, an efficient 
multimodal deep learning module is used to estimate the clean speech(es), which
can be used as clean speech or passed through a speech-to-text engine to
convert into commands. 

\begin{figure}
	\centering
	\includegraphics[width=1\columnwidth]{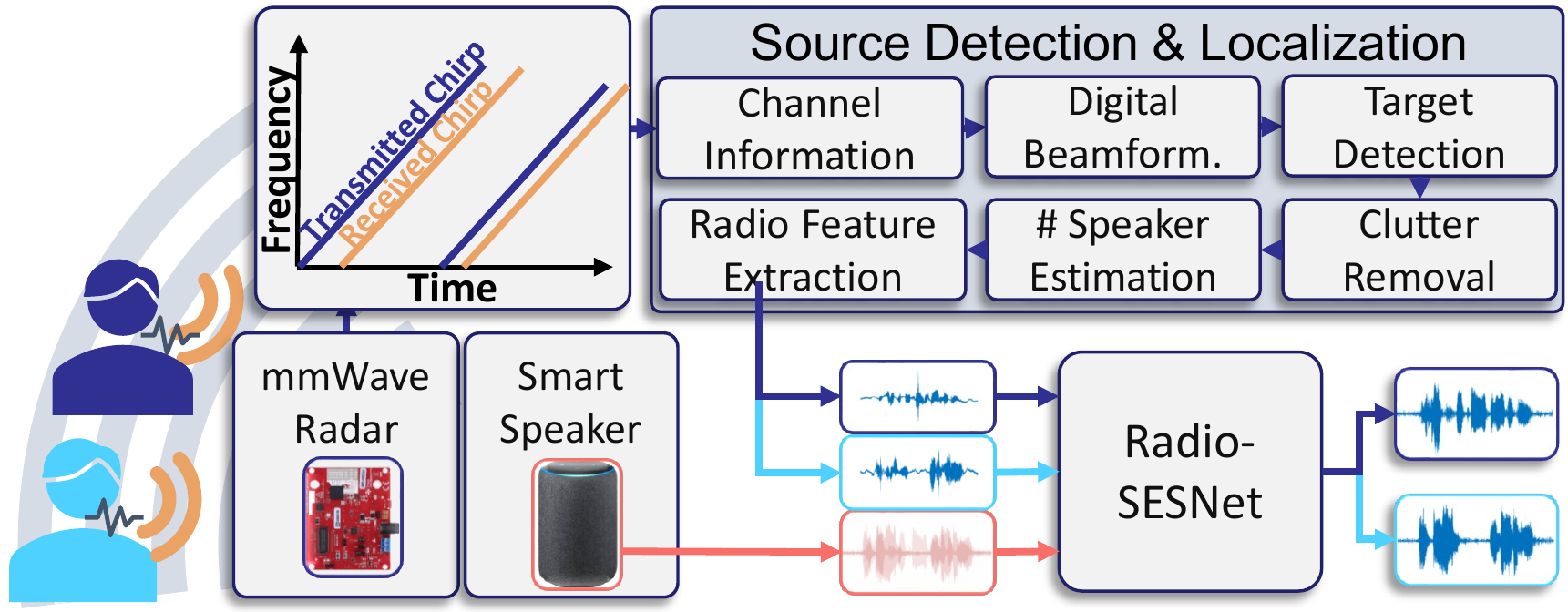}
	\caption{\sysname Design}
	\label{fig:radiosesdesign}
\end{figure}

The first main block of \sysname, source detection and localization in Fig.
\ref{fig:radiosesdesign}, is explained briefly in
\S\ref{sec:radio-preprocessing}, whereas the second block, deep learning module 
is
further detailed in \S\ref{sec:deeplearning}.

  \begin{figure*}[!ht]
    \subfloat[
		2D CFAR window
		\label{subfig:preprocess-cfar}]
		{%
      \includegraphics[width=\subfigwidth\textwidth]{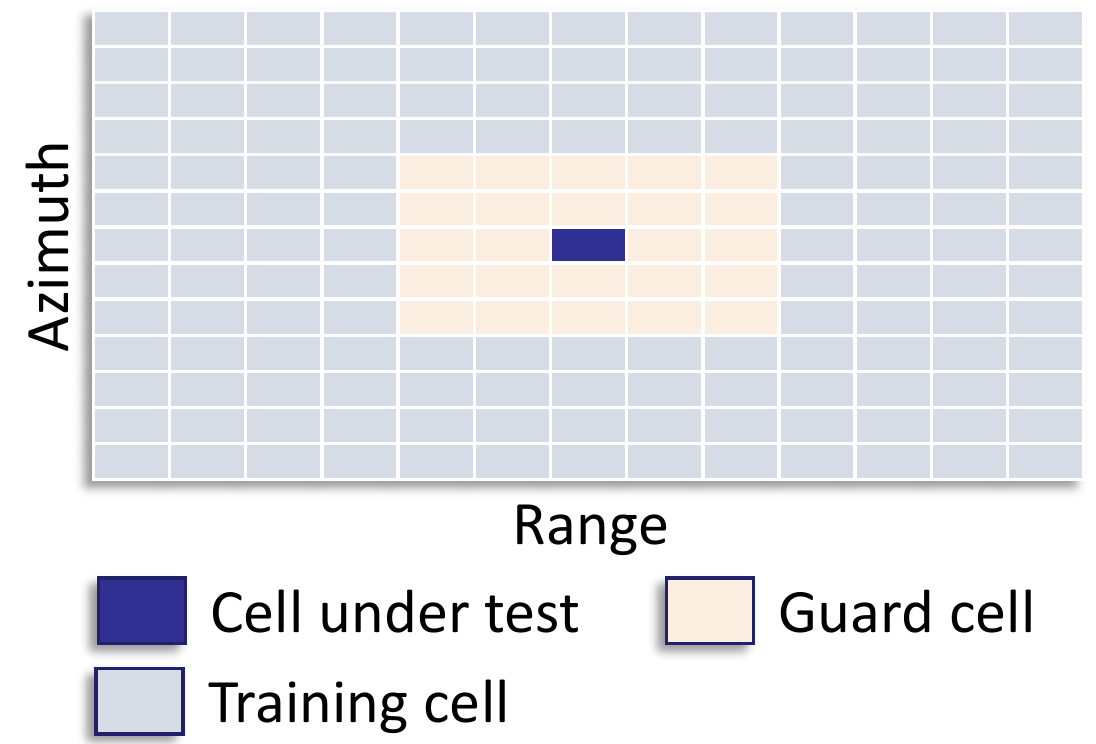}
    }
    \hfill
    \subfloat[Amplitude map $|h_{r,\theta}(t)|$\label{subfig:preprocess-amplitude}]{%
      \includegraphics[width=\subfigwidth\textwidth]{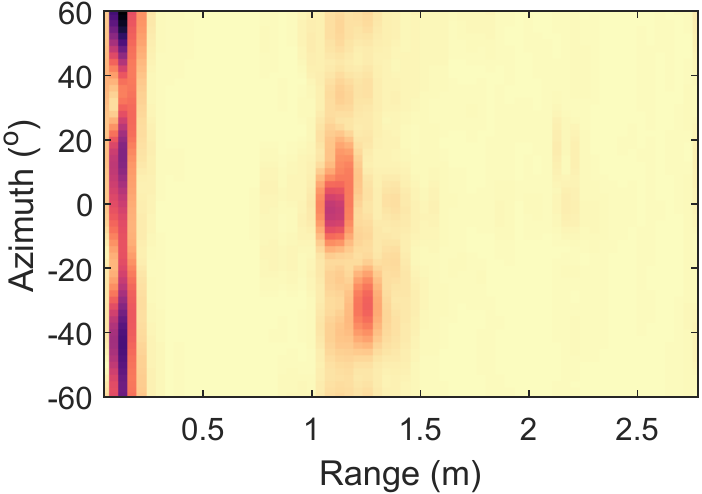}
    }
	\hfill
    \subfloat[Variance map ($V_{r,\theta}(t))$\label{subfig:preprocess-variance}]{%
      \includegraphics[width=\subfigwidth\textwidth]{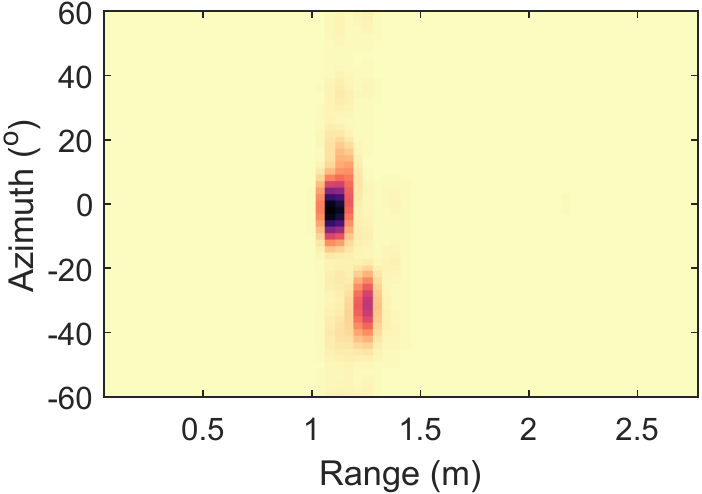}
    }
	\hfill
    \subfloat[Detection map ($B_{r,\theta}(t)$)\label{subfig:preprocess-detection}]{%
      \includegraphics[width=\subfigwidth\textwidth]{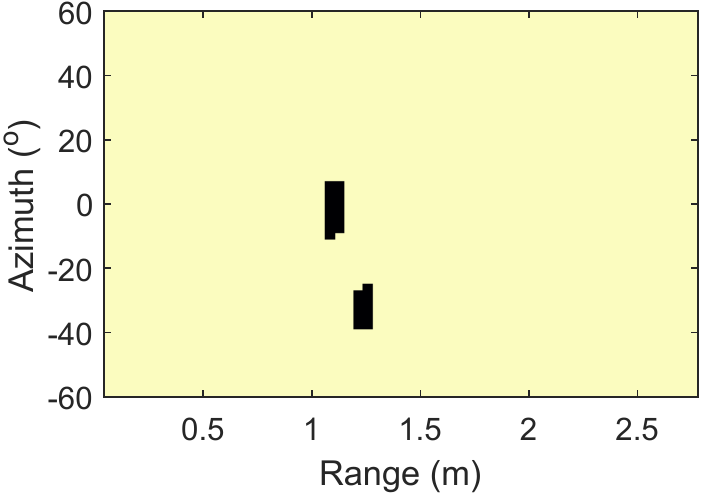}
    }
	\hfill
    \subfloat[Clustering output\label{subfig:preprocess-clustering}]{%
      \includegraphics[width=\subfigwidth\textwidth]{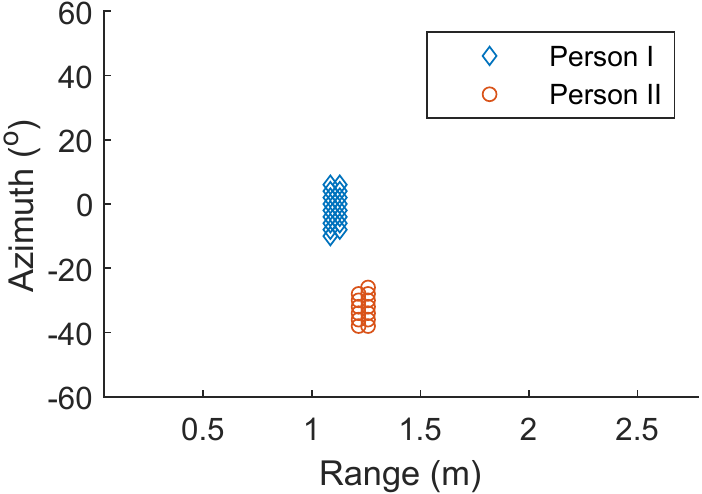}
    }
    \label{fig:dummy}
    \caption{Illustration of Sound Detection \& Localization Module of \sysname}
  \end{figure*}

\section{Radio Feature Extraction}
\label[]{sec:radio-preprocessing}

As shown in Fig. \ref{fig:radiosesdesign}, the goal of the radio feature
extraction module is 
to output individual radio streams from sources in the
environment. 
To achieve that, we adapt a variety of methods in an efficient pipeline to detect and locate targets. 
Unlike existing works, such as \cite{ozturk2021radiomic,uwhear},
\sysname does not rely on a spectrogram-based metric to localize people in the 
environment, but utilizes classical, efficient methods to extract the corresponding 
range-azimuth bins. 

\head{Channel Information}
\label[]{subsec:radio-1-channel}
\sysname can work with any type of radar that can report a \textit{channel impulse
response} (CIR), although we use a frequency modulated
continuous wave (FMCW) radar. When using an FMCW radar, extracting the
CIR requires applying an operation called
range-FFT, which is a common operation and we refer the reader to related work
\cite{stove1992linear}. As mmWave devices usually have multiple
antennas, we define the CIR at the $m$-th antenna
$h_m(\tau)$ as:
\begin{equation}
	h_m(\tau) = \sum_{r=0}^{R-1}\alpha_{m,r}\delta(\tau-\tau_r) + \epsilon(\tau),
	\label[]{eq:cir}
\end{equation}
where $R$ is the number of the CIR range bins, $\delta(\cdot)$ is the Delta
function representing the presence of an object at the corresponding location,
$\alpha_{m,r}$ and $\tau_r$ denote the complex amplitude and the propagation delay of
the $r$-th range bin, and $\epsilon$ denotes the additive noise, respectively.
Here, the range
resolution $\Delta R$ can be inferred from the time resolution, $\Delta \tau$, 
which is inversely proportional to bandwidth (corresponding to $4.26cm$ for
our device). Therefore, a separate stream from very close targets can be
extracted.
The CIR in \eqref{eq:cir} is captured repeatedly during sensing, and is time
dependent. To simplify \eqref{eq:cir}, we denote the CIR from $m$-th antenna, at
$r$-th range bin, at time index $t$ as $h_{m,r}(t)$. Note that, $h_{m,r}(t)$ is
quantized with respect to time, range bin, and antenna index.

\head{Digital Beamforming}
\label[]{subsec:radio-2-bfr}
Using the individual received streams from each antenna, \sysname extracts
range-azimuth information with classical beamforming \cite{van1988beamforming}.
Range-azimuth CIR is denoted by $h_{r,\theta}(t)$, where $\theta$ represents the
azimuth angle. Since our virtual antenna array elements are placed $d=\lambda/2$
apart, where $\lambda$ is the wavelength, $h_{r,\theta}(t)$ can be given as:
\begin{equation}
	h_{r,\theta}(t) = \mathbf{s}^{H}(\theta)\mathbf{h_{m,r}(t)} + \epsilon(t),
\end{equation}
where $\mathbf{s}^H(\theta)$ is the steering vector for angle $\theta$, and
$\epsilon$ is the additive noise. The coefficients of the steering vector are:
\begin{equation}
	s_{m}(\theta) = \exp \left(-j2\pi\frac{d \sin \theta}{\lambda}\right),
\end{equation}
and the channel vector is $\mathbf{h_{m,r}(t)}= [h_{1,r}(t), h_{2,r}(t), ...,
h_{M,r}(t)]$, with $M$ being the total number of antenna elements.

\head{Target Detection}
\label[]{subsec:radio-3-target}
To detect human bodies in the environment, \sysname first
extracts the reflecting objects in the environment. As suggested by
\eqref{eq:cir}, the presence of objects creates strong returned signals, whereas
when there is no object, returned signals only consist of noise.
For target detection, we utilize a classical approach in the radar literature,
constant false alarm rate (CFAR) detector \cite{richards2005fundamentals}, which
adaptively estimates the background noise for different bins and
thresholds each range-azimuth bin accordingly. As shown in
Fig. \ref{subfig:preprocess-cfar}, the 2D CFAR window is denoted
with
$C$, and CFAR threshold is denoted with
$\gamma$. This window is applied to the magnitude of the range-azimuth plane, 
and the corresponding range-azimuth
plane is shown in Fig. \ref{subfig:preprocess-amplitude}.
Therefore, the CFAR detection rule on the range-azimuth plane is given as:
\begin{equation}
	B^{\text{CFAR}}_{r,\theta}(t) = \mathds{1}\{(C \star |h_{r,\theta}|)(t) > \gamma(|h_{r,\theta}(t)|) \},
\end{equation}
where $\star$ and $\mathds{1}\{\cdot\}$ denote the convolution operation and indicator
function, respectively.

\head{Clutter Removal}
\label[]{subsec:radio-4-clutter}
Previous module extracts a binary map with bins with reflecting
objects, which can include static objects.
On the other hand, even when a person is stationary, the
radar signal still captures a variation at the person's location, due to
inherent body motion from breathing and heart rate, a phenomenon used
extensively in mmWave based person detection \cite{wang2021mmhrv,adib2015multi}.
Therefore, to remove the static objects and detect human bodies, we
extract the variance at
each range-azimuth bin, and use a threshold to identify static objects. We
denote 
the variance of $h_{r,\theta}(t)$ with $V_{r,\theta}(t)$, where an example can
be seen in Fig. \ref{subfig:preprocess-variance}. Therefore, human
detector output is $B^{\text{stat}}_{r,\theta} \triangleq \mathds{1}\{V_{r,\theta}(t) >
\mathrm{H}^{\text{stat}}(r,\theta) \}$. Furthermore, bodies with excessive motion can also be
filtered using a similar approach, and we reject those by: $B^{\text{mov}}_{r,\theta}
\triangleq \mathds{1}\{V_{r,\theta}(t) < \mathrm{H}^{\text{mov}}(r,\theta) \}$,
where $\mathrm{H}^{\text{stat}}(r,\theta) \triangleq \frac{\eta^\text{stat} \cos
(\theta)}{(1+r\Delta R)^2}$, $\mathrm{H}^{\text{mov}}(r,\theta) \triangleq
\frac{\eta^\text{mov} \cos (\theta)}{(1+r\Delta R)^2}$, $\eta^\text{stat}$ and
$\eta^\text{mov}$ are empricially found thresholds. The minimum and maximum
variances
are defined with respect to $(r,\theta)$, in order to accommodate changing
reflection energy with respect to angle and distance. The resulting binary
detection map, $B_{r,\theta}(t)$
is found by extracting intersection of all binary maps, i.e. $B_{r,\theta}(t)=
\{ B^{\text{CFAR}}_{r,\theta} \cap B^{\text{stat}}_{r,\theta} \cap
B^{\text{mov}}_{r,\theta}\}(t)$, as shown in Fig. \ref{subfig:preprocess-detection}.

\head{Number of People Estimation}
\label[]{subsec:radio-5-numpeople}
Each bin of binary detection map, $B_{r,\theta}(t)$ spans $(\Delta R, \Delta
\theta)$ distance in 2D space. Considering the high range and angular
resolution, 
a human body can span multiple bins in $B(r,\theta)$. To estimate the number of 
people, \sysname clusters binary detection
maps using a non-parametric clustering method, DBSCAN \cite{ester1996dbscan}.
The parameters for DBSCAN are set empirically, and an example clustering is
shown in Fig. \ref{subfig:preprocess-clustering}. Furthermore, since 
the number
of people estimation and center extraction is done repeatedly for a window of
size $W$, there is a need to match the locations of bodies at different time
indices. We use Munkres' algorithm \cite{munkres1957algorithms} to continuously
track the location of users. 

\head{Radio Feature Extraction}
\label[]{subsec:radio-6-feature}
Having extracted the number of persons and the corresponding range-azimuth bins,
\sysname extracts the complex radar signals from each person's
center directly, following recent raw-data based approaches
\cite{zheng2021morefi}.
As there are many range-azimuth bins associated with the same person, \sysname
extracts the median bin for testing, whereas multiple nearby bins are used for
training, which helps to boost dataset size and mitigate overfitting. 
Output dimensionality of the
radar signals is $2\times 1000$ at 16bits for a 1-second stream, 
which is lower than the
microphone and typical video streams. 

\vspace{-0.1in}
\section{AudioRadio Deep Learning Model}
\label[]{sec:deeplearning}
In this section, we explain the structure of the deep learning model used in
\sysname, named \dlmodule. We first introduce the relevant background in SES
and our design rationale in
\S\ref{subsec:background} and then detail our design in \S\ref{subsec:designdlmodule}.

\subsection{Background and Design Rationale}
\label[]{subsec:background}

\head{Background:}
Usually, an SES model follows the
architecture in Fig. \ref{fig:typicalsystem}, with an encoder, masker,
and a decoder block \cite{michelsanti2021overview}. 
Input encoding is multiplied with an estimated mask, which
uses a decoder to reconstruct the time-domain signal.
Early works have used STFT as the encoder, with the ideal binary mask
being the training objective \cite{roman2001speech}. The performance can be
increased by using more optimal masks (such as complex ratio mask
\cite{williamson2016complex}); however, these still suffer from the fact that 
STFT-based encoding is not necessarily
optimal for speech separation, and methods that replace STFT
with adaptive encoders are found to be more optimal \cite{luo2019conv}.

\head{Design Rationale:}
\sysname uses the same structure as in Fig. \ref{fig:typicalsystem}, with the
addition of a radio stream. 
Radio streams are encoded, and concatenated with the audio stream to
estimate the masks. However, this involves a few design choices as follows:
Unlike audio signals, radio signals are complex-valued, and both real
and imaginary parts change with respect to the motion and vibration
\cite{zheng2021morefi, jiang2020mmvib}.  
If a spectrogram representation is used as an input, 
not only it may not be optimal
for neural network, but it usually involves in throwing away some signal content
by only extracting amplitude, or 
half of the spectrogram (\eg~only positive Doppler shifts), 
as in \cite{xu2019waveear,
liu2021wavoice}. Using either the real or imaginary part of 
the signal (as
in \cite{uwhear}) or combining both parts optimally with a linear projection
\cite{ozturk2021radiomic} also loses important signal content.
Based on this, \sysname uses adaptive front-end for radio streams.

To make \sysname work with raw radio inputs, we apply random rotation
in IQ plane, as proposed by previous work \cite{zheng2021morefi}\footnote{We
refer the reader to \cite{jiang2020mmvib} for IQ representation of the returned
signals, and to \cite{zheng2021morefi} for discussion and introducing random rotation.}. 
However,
unlike \cite{zheng2021morefi}, we apply a high-pass filter on
returned signals to reduce the effect of
body motion. The high pass filter is needed for \dlmodule to run with
raw radar inputs, as will be shown in \S\ref{sec:evaluation}.
We select the cutoff frequency of the high pass filter at 90 Hz in order not to
filter vocal folds harmonics. Afterward, the radio signals are encoded
with an adaptive encoder, as 
explained in \S\ref{subsec:designdlmodule}.

After the encoder, we process
audio and radio streams separately with individual blocks to exploit long-term dependencies
within each modality. 
To that end, we process each modality through an efficient dual-path RNN
block (DPRNN). 
DPRNN blocks do not suffer from limited context, a main issue with
fully convolutional models \cite{luo2020dual}. Afterward, 
we combine two modalities via resizing and concatenation on the
feature dimension. These models are further processed with DPRNN blocks and 1D
decoders before outputs. 

\begin{figure}
	\centering
	\includegraphics[width=1\columnwidth]{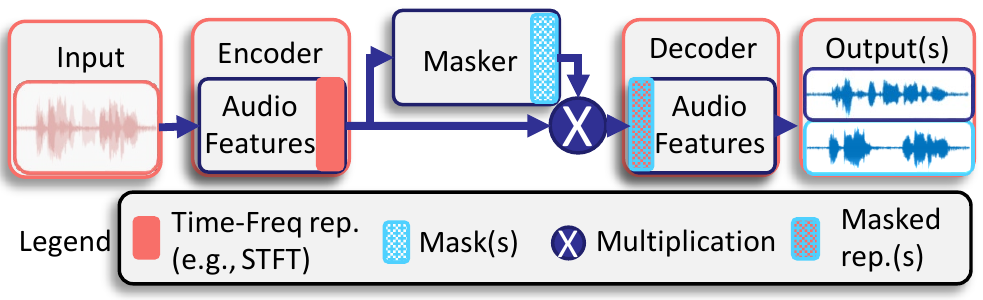}
	\caption{Typical SES System Workflow}
	\label{fig:typicalsystem}
\end{figure}

\subsection{\dlmodule Design}
\label[]{subsec:designdlmodule}

\subsubsection{Encoders}
\label[]{subsec:encoder}
The audio
encoder of \dlmodule consists of a 1D convolutional layer, with kernel size 16, 
and number of kernels
256, followed by ReLU nonlinearity and layer normalization. Radio channel uses
another 1D convolutional layer, nonlinearity and normalization, with the same
parameters, except the number of filters being 64, due to the lower sampling rate.
Stride size is set to 1/2 of the kernel width, resulting in $50\%$ overlap
between convolutional blocks. After the first layer, a second 1D convolution
reduces the dimensionality to 64 for audio, and 16 for radio.
Each radio stream uses the
same encoder block to create an STFT-like representation. 
We denote the distorted input audio with $\tilde{a}$, and radio streams with $r_i$, where $i$
denotes the $i^{th}$ radio stream. Output of the audio and radio encoders are 
represented with $\mathbf{X_{{\star}}} \in
\mathcal{R}^{N_{{\star}} \times L_{{\star}}}$, with $\star \in (a,r)$,
for audio and radio stream, where we drop the index $i$ for simplicity. Here,
$N_{\star}$ represents the number of features, and $L_{\star}$
represents the number of time samples of encoded representation.

\begin{figure*}[t]
	\centering
	\includegraphics[width=0.8\textwidth]{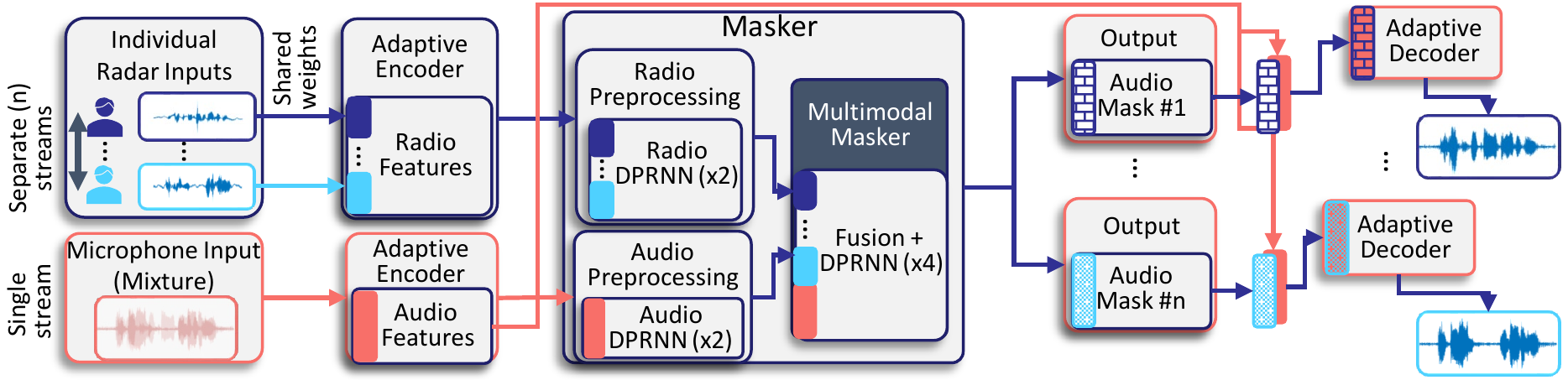}
	\caption{\dlmodule Structure}
	\label{fig:neuralnetstructure}
\end{figure*}

\begin{figure}[t]
	\centering
	\includegraphics[width=1.0\columnwidth]{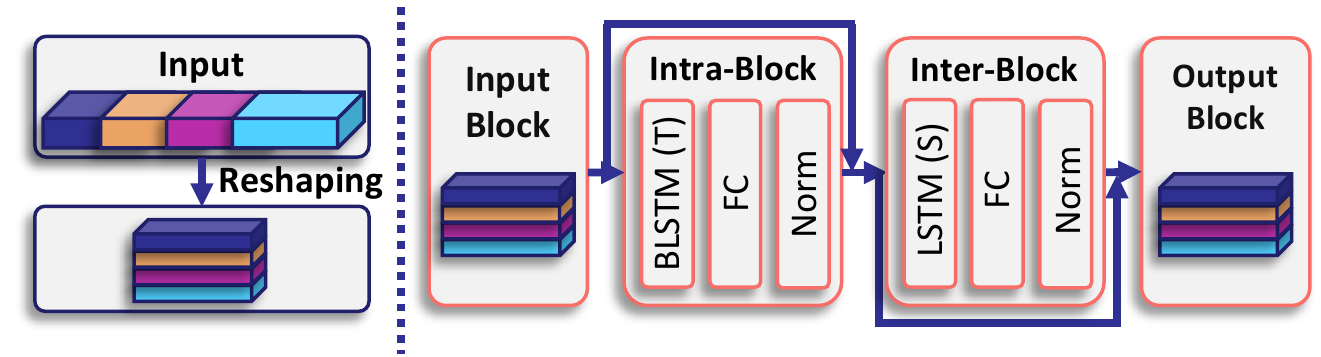}
	\caption{Left: Reshaping operation with overlapping
	windows. 
	Right: Single DPRNN Block}
	\label[]{fig:dprnn-block}
\end{figure}

\begin{table}[b]
	\caption{Parameters for the Masker Layer for 2-Mix} 
	\label{tab:masker-params}  
	\centering      
	\begin{tabular}{l | c  c | c  c | c  c }
		\hlineB{4}
		Audio & $N_{\text{a}}$ & 64 & $K_{\text{a}}$ & 128 & $S_{\text{a}}$ & 48 \\
		Radio & $N_{\text{r}}$ & 16 & $K_{\text{r}}$ & 16 & $S_{\text{r}}$ & 48
		\\
		Concatenation & $N_{\text{c}}$ & 96 & $K_{\text{c}}$ & 128 & $K_{\text{c}}$ & 48 \\   
		\hlineB{4}
	\end{tabular}
\end{table}

\subsubsection{Masker}
\label[]{subsec:masker}
Both encoded modalities are combined to estimate the masks for
each source, as illustrated in the masker of Fig. \ref{fig:neuralnetstructure}.
Each modality passes through individual
DPRNN blocks, then fused by vector concatenation, and
passes through four more DPRNN blocks before estimating the mask with a 2D
convolutional layer, which matches the output with the expected mask number and
size.

\head{DPRNN Processing:}
For processing the encoded data, we use DPRNN blocks \cite{luo2020dual}, where an
example DPRNN workflow is presented in Fig. \ref{fig:dprnn-block}. DPRNN
processing 
consists of reshaping the input data to a 3D representation, through means of
extracting overlapping blocks, and concatenating through another dimension,
and
applying two consecutive RNN layers to different dimensions of the input
block. The output of
the reshaping operation can be
represented as $\mathbf{\hat{X}_{\text{a}}} \in
\mathcal{R}^{N_{\text{a}} \times K_{\text{a}} \times S_{\text{a}}}$, with
$K_{\text{a}}$ and $S_{\text{a}}$ denoting the block length and number of
blocks. 
The input, output representations $\mathbf{X_{\text{r}}}$,
$\mathbf{\hat{X}_{\text{r}}}$ and dimensionalities $N_{\text{r}}$,
$L_{\text{r}}$, $K_{\text{r}}$ and $S_{\text{r}}$ are defined similarly for radio
channel, and given in Table \ref{tab:masker-params}, whereas the flow for a
single DPRNN processing is given in Fig. \ref{fig:dprnn-block}.

After a suitable reshaping operation, the input blocks are fed to an RNN module,
which is operated along the $S$ dimension of the 3D input, followed by a fully
connected layer, and layer normalization. After a skip connection in between, 
a similar operation is 
repeated through
$K$ dimension to capture larger distance relationships between blocks. Each RNN
block has depth 1, and fully connected layers are used to match
the input size to the output size, which enables to repeat multiple DPRNN blocks
without any size mismatches.

\subsubsection{Decoder}
\label[]{subsec:decoder}
At the output of the masker, a number of masks equal to the number of people are
estimated, which is then used to \textit{decode} the signal to extract time
domain audio signals. DPRNN blocks are converted
back to a representation similar to the one at the input, 
by overlap-add method \cite{luo2020dual}. The signal is fed through
the decoder, which applies a transposed convolution operation.
The output is a single channel representation, 
with the same dimensionality and the same number of filters in the
encoder to preserve symmetry, and it is also adaptive.

\subsubsection{Training}
\label[]{subsec:training}
In order to train \dlmodule, we use scale-invariant
signal-to-distortion (SiSDR \cite{le2019sdr}) as the loss function
between the time-domain signals, which is given by:
\begin{equation}
	\mathbf{SiSDR(a, \hat{a})} = 10 \log_{10} \left(\frac{||\frac{\hat{a}^Ta}{||a||^2}a||}{||\frac{\hat{a}^Ta}{||a||^2}a -\hat{a}||}\right),
\end{equation}
where $a$ and $\hat{a}$ denote the target and the estimated sound signals. Use
of SiSDR prevents scaling effects to dominate the error calculation, as the
amplitude of extracted speech is not of interest.
The SiSDR loss has been combined with $L_2$ norm regularization on the weights,
where the decay factor is set to $1e^{-6}$.  
Since a separate model for different numbers of users
has been trained,  
\sysname switches to the appropriate model by estimating the number of sources.

\subsubsection{Other Design Considerations:} 
\label[]{subsubsec:other} Complexity and causality are particularly considered in our design.

\head{Complexity:} \dlmodule has a compact design, with only 2.1M
parameters. Among these, radio stream occupies 320k parameters, 
which could easily be fit on a small device. Forward
pass of a 3-second input with \dlmodule takes 
4ms on a modern GPU with batch processing,
which is only 0.4ms slower than the corresponding audio-only method. 

\head{Causality:} \dlmodule uses unidirectional LSTMs in the recurrent layers of
inter-block processing, whereas intra-blocks rely on BLSTMs which requires having the complete block in $S$ dimension. Therefore,
\dlmodule can work in a causal fashion, with roughly $150ms$ delay. We leave
investigation of a real-time work to future, but \sysname is already close to
real-time processing, unlike \cite{sun2021ultrase,ephrat2018looking}.

\vspace{-0.1in}
\section{Experiment and Implementation}
\label[]{sec:implementation}

\subsection{Data Collection}
\label[]{subsec:dataset}
\head{Hardware:}
We build a data collection platform, as seen in Fig. \ref{fig:datasetup}, to
obtain large-scale data to train, validate, and evaluate \sysname. As extracting
clean and non-reverberant
ground truth samples are important, we reduce the echo in the room by
sound-absorbing pads. 
We collect clean audio data with a Blue Snowball 
iCE microphone, sampled at 48 kHz, radar data using a Texas Instruments (TI) IWR1443 
mmWave
radar, and video data using the front-facing camera of an iPhone 11 Pro. 
The radar is set to operate with a
bandwidth of 3.52 GHz at a sampling rate of 1000 Hz. 
We align
the radio signal and audio signal in the time domain using the correlation of
their energy. 
Video data, captured at 1080p and 30 fps, is collected for future research and
not used in this work; although the accompanying audio files are used for training.

\begin{figure}
	\centering
	\includegraphics[width=0.6\columnwidth]{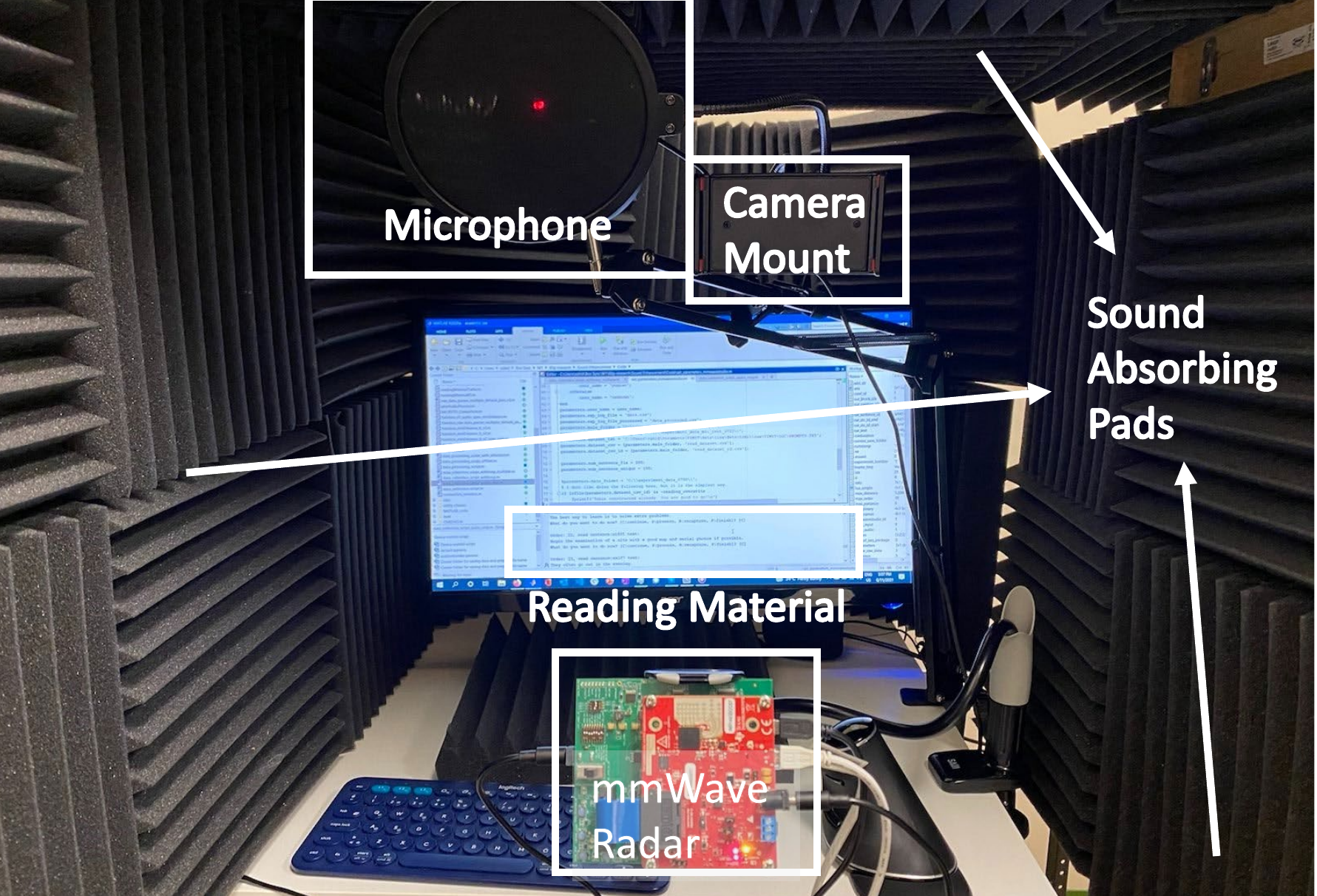}
	\caption{Setup of Data Collection Center}
	\label{fig:datasetup}
\end{figure}

\head{Speaker setting:}
We recruit 19 users including native speakers and speakers with
different accents to read phonetically rich sentences from the TIMIT 
corpus \cite{garofolo1993timit}. Our speakers come from a diverse background,
where
there are 5 native English speakers, along with 9 Chinese, 2
Indian, 2 Turkish, 1 Korean accented speakers.
We remove
sentences that are shorter than 25 characters in the dataset. Since the size of
TIMIT corpus is limited, 200 common and 100 unique sentences are read by each
participant. A total of 5700 sentences were read, including 2100
\textbf{unique} sentences and 5762 \textbf{unique} words. The sentences are
presented in mixed order, and
our dataset includes a lot of pauses and filler words, in contrast to publicly
available datasets, which usually include professional speakers 
(\eg ~LibriMix \cite{cosentino2020librimix}). 
During data collection, users sit approximately 40cm away from
the radio device and read each material at a normal speaking volume while
not moving excessively.

\head{Data generation:}
To generate the noisy and mixture sound signals, we follow the recipe used in
LibriMix \cite{cosentino2020librimix} with the noise files from WHAM dataset
\cite{wichern2019wham}.
We randomly select 13 users for training, and 4 users (2 male,
2 female) for evaluation. Validation set includes the remaining two users, and
unused speech of the users in the 
training set.
After downsampling all audio files to 8kHz, we create
synthetic mixtures based on the shortest of the combined files, 
with a minimum duration constraint of 3-seconds.
Each
user's recordings are repeated ten times on average, which results
in 25,826 utterances ($\approx$30 hours).
The gain factors are 
found by normalizing the loudness of speech and noise signals, and creating
noisy mixtures in $[-5,5]$ dB 
signal-to-noise rate (as in \cite{cosentino2020librimix}).
We create two evaluation sets: 
\begin{itemize}
    \item \textbf{Seen:} mixtures from seen users, but unheard sentences (\aka~
    closed-condition)
    \item \textbf{Unseen:} mixtures from
    unseen users (\aka~open-condition)
\end{itemize}
This helps us to better understand the dependency on seen/unseen users
in \sysname, as different users' radio signals can be different, not only due to
their speaking, but also due to their body motion and physical characteristics.
Other experimental settings are also
introduced and investigated in \S\ref{sec:evaluation}. On the other hand, our
experiments include mostly overlapping speech, to better illustrate the
difference between audio-only and audioradio methods, and we leave evaluation of
partially overlapping speech to future for conciseness.

\head{Dataset Considerations for Improving Robustness:}
A multimodal system can fail easily and focus to use a single modality, which is
known as mode failure. To prevent this and to further improve robustness, our
dataset creation procedure includes the following:
\begin{itemize}
	\item \head{Same-speaker mixtures:} 
	Our dataset includes
	same-speaker mixtures, in order to prevent mode failure,
	which is shown to be effective in the audiovisual domain \cite{gabbay2018visual}.
	\item \head{Multi-microphone mixtures:} As our data collection procedure
	includes two microphones, we randomly select one when generating
	each mixture. Our evaluation is done with the
	better microphone (Blue), but this also boosts dataset size multiple folds
	without collecting more data.
	\item \head{Clean and Noisy Mixtures:} Unlike the LibriMix dataset
	\cite{cosentino2020librimix}, we create both noisy and clean mixtures of
	multiple speakers and use them to train a single model. Therefore, \sysname
	uses a single model, whether an environment is clean or noisy.
\end{itemize}

\vspace{-0.1in}
\subsection{Implementation Details}
\label[]{subsec:implementation}
We implement data collection and raw data processing modules of \sysname in
MATLAB, whereas the deep learning model is implemented in PyTorch, with the help
of Asteroid library \cite{pariente2020asteroid} to follow standard training
and evaluation protocols in monoaural SES, and to borrow implementations of 
existing methods, such as ConvTasNet \cite{luo2019conv} or DPRNNTasNet
\cite{luo2020dual}. 
We train \dlmodule
and DPRNNTasNet for 60 epochs, using a starting learning rate of $1e^{-3}$, which is
halved when the validation loss did not improve for 5 consecutive epochs.
Furthermore, the learning rate is scaled by $0.98$ every two epochs, as in
\cite{luo2020dual}. 
An early stopping criterion is set to 15 epochs. 
To accelerate training, we use
mixed-precision training. 
Thanks to the low complexity design of \dlmodule, 
a single epoch takes roughly 10 minutes to train, with a batch size of 24, 
using a single NVIDIA RTX 2080S
GPU.

\head{Considerations to Improve Robustness:}
As noted previously, although microphone signals
mostly correspond to speech signals, radar signals can be affected by motion, 
vibration, and environmental factors. Furthermore, it is usually not
straightforward to make a multimodal system work easily. To improve the
robustness of radio signals, we implement the following:
\begin{itemize}
	\item \head{Capturing Multiple Snapshots} Since a single user spans multiple
	range-azimuth
	bins due to high resolution, we record multiple range-azimuth data in our dataset. In each
	epoch, we randomly select a range-azimuth bin for training among a maximum
	of 8
	candidates, 
	whereas
	validation and testing use the median bin. This boosts the
	dataset size significantly without relying on synthetic methods
	and enables to use a wider range of bins, instead of searching
	for the most optimal bin. 
	\item \head{Input Distortions:} The input radio streams are distorted in
	different ways. These include introducing random rotation \cite{zheng2021morefi}, 
	adding noise at
	different variance levels, replacing some part of the radio signal with zeros 
	(to
	imitate data loss), or removing some radio signals
	completely, as suggested by \cite{ngiam2011multimodal} to reduce mode
	failure.
\end{itemize}

\section{Evalution}
\label[]{sec:evaluation}
In this section, we introduce performance metrics, and baselines for comparison,
which are followed by results using \sysname.  
Afterward, we investigate the practical limits and robustness of \sysname
by analyzing environmental effects.
Next, we present a real-world case study to illustrate the benefits coming from \sysname. Last, we
evaluate \sysname in some interesting cases, such as noisy, partial inputs, and
conduct an ablation study.

\head{Performance Metrics:}
We report the following metrics to evaluate performance of \sysname:
\begin{itemize}
	\item SiSDR \cite{le2019sdr}: Scale-invariant signal-to-noise ratio, which
	is an indicator of signal levels, with a normalization factor to prevent
	scaling of the signals to increase metric unfairly.
	\item SIR: \cite{vincent2006performance}: Signal-to-interference ratio, which measures the leakage from one
	person to another when there are multiple speakers, and only reported for
	SS tasks.
	\item STOI \cite{taal2011shorttime}: Short time intelligibility metric,
	correlates with the word error rate, reported from 0 to 1.
	\item PESQ \cite{rix2002perceptual}: Perceptual evaluation of the sound
	quality, measured from 0 to 5. Since measuring human perception requires user studies, this metric has been proposed as an alternative, when user studies are not feasible. 
\end{itemize}

\head{Baseline Methods:} We include several radio-only and
audio-only methods in the literature for a variety of tasks. First, as a
radio-only method, we implement WaveVoiceNet in WaveEar \cite{xu2019waveear}.
This approach uses the radio
modality alone to (re)construct sound signals from vocal folds vibration, 
and assumes no available
microphones. It reconstructs magnitude of 
audio spectrograms and uses Griffin-Lim
based phase reconstruction. We use oracle phase of the clean audio signal
instead, 
which poses
an upper limit on its performance. 
Another recent work \cite{liu2021wavoice} is similar to our work in
combining two modalities, yet their end-to-end system focuses on translating
single speaker noisy voice commands into text without a sound output and not
comparable 
to our method. 

We compare performance
of \sysname with other audio-only baselines, to illustrate gains from radio
modality, and sustained performance of \sysname. We include ConvTasNet
\cite{luo2019conv}, one of the first adaptive-encoder based systems that
outperformed STFT-based masks. Second, we include DPRNNTasNet, which is the
audio-only baseline of \sysname. DPRNNTasNet has shown to
outperform ConvTasNet significantly, and can be considered as the state-of-the
art. 
Third, 
we use SudoRMRF \cite{tzinis2020sudo}, which simplifies DPRNNTasNet by replacing 
the RNN blocks with
downsampling and upsampling blocks and is shown to achieve similar performance.

Last, we cannot compare with UltraSE
\cite{sun2021ultrase},
as it uses ultrasound
modality, and different speakers and noise dataset. Due
to changes in datasets and different sampling rate (16 kHz), it is not possible
to copy their results and draw a direct comparison. On the other hand, UltraSE
performs similar to ConvTasNet in 2-person mixtures, which we have included as a
benchmark in our study.

\begin{table}[t]
	\caption{Results for enhancing single speaker speech. Seen: 
	closed-condition, and unseen: open-condition} 
	\label{tab:enh-single}  
	\centering     
	\begin{tabular}{ p{25mm} || c | c | c || c | c | c}
		\hlineB{4}  
		Evaluation & \multicolumn{3}{c ||}{Seen} & \multicolumn{3}{c }{Unseen}
		\\ 
		\hline
		Model & \metricone &
		\metricthree & \metricfour 
		& \metricone & \metricthree & \metricfour	\\
		\hline
		Input        	& 3.9  	       & 0.74  		& 1.55 	    &	3.8  	    & 0.70  	 & 1.54 	 \\ 
		WaveVoiceNet 	& 0.6          & 0.60       & 1.28  	&	0.7         &  0.62      & 1.27      \\
		ConvTasNet    	& \bd{14.5}    & 0.90 		& 2.67     	&	\bd{13.6}   & \bd{0.87}	 & \bd{2.55} \\
		SudoRMRF       	& 14.0         & 0.88		& 2.32     	&	12.2  	    & 0.84	     & 2.04      \\
		DPRNNTasNet     & 14.2         & 0.89		& 2.62    	&	13.0        & 0.86	     & 2.46	     \\
		\sysname     	& 14.5         & \bd{0.90 } & \bd{2.68} &	13.3        & 0.87       & 2.52 \\
		\hlineB{4}
	\end{tabular}
\end{table}

\vspace{-0.05in}
\subsection{Speech Enhancement}
\label[]{subsec:perf-overall-enh}
In speech enhancement, \sysname brings improvements to the audio-only baseline
methods, as shown in Table \ref{tab:enh-single}. Since the background signals are
statistically different than speech signals, we see relatively small
improvements. This observation is consistent with
audiovisual methods (\eg~0.1 dB improvement in \cite{ephrat2018looking}), and
shows that \sysname learns to exploit the radio information. 
On the other hand, results from WaveVoiceNet suggest that, radio modality is not
sufficient to (re)construct less-noisy audio, and may not be feasible within our
experimental setting. This can be attributed to differences in the hardware
(special hardware is used in \cite{xu2019waveear}), our phonetically rich 
diverse dataset 
(5762 unique words vs. 631 in \cite{xu2019waveear}), and users. As the results
are poor, we do not investigate WaveVoiceNet further in our experiments.
Performance of \sysname is matches to that of ConvTasNet, with certain
qualitative differences, such as 1.5s look-ahead in ConvTasNet, and higher
computational complexity. We also note that, our
implementation uses a pretrained ConvTasNet on a much larger dataset, which
potentially improves the overall performance.
This section investigates the case, where the background is non-speech noise.
Having an interfering speech signal can also be considered as speech enhancement
problem, yet the enhancement methods usually require some prior information to
focus on the particular speech. If such prior information does not exist, it is
more reasonable to evalutae the performance against speech separation models. In
order to have a fair comparison, we evaluate
this case in the following sections, under speech separation. 

\subsection{Speech Separation}
\label[]{subsec:perf-overall-sep}
In this section, we present the speech-separation results with \sysname, 
along with the previously
mentioned baselines in 
Table \ref{tab:sep2}.
For both separating single and noisy speech
tasks, \sysname outperforms a variety of state-of-the-art methods in audio-only
domain, including DPRNNTasNet. Our DPRNNTasNet implementation achieves 13.5
SiSDR in 2-person clean mixtures, which is
close to the reported value in the LibriMix dataset, 16.0.
Significant improvements with respect to SIR
can be observed in both clean and noisy cases,
which indicates the usefulness of radio channel for separating the mixtures, and
suppressing the interference.
Furthermore, 
even though there is more variety in radio inputs (\eg~radio channel inputs are
not only affected by the sound, but also by ambient motion and physical
characteristics), 
\sysname can
still generalize better to unseen users, where the basic DPRNNTasNet suffers.
\sysname not only improves signal metrics, but also intelligibility and
the perceptual quality metrics (PESQ). The difference between the audio-only
baseline becomes larger, especially when the input mixtures are corrupted with
noise and when there are multiple people. We also train \sysname
with three people mixtures. As
shown in Table \ref{tab:sep3}, the improvements from \sysname is even greater
for 3-person
mixtures, as radio helps to extract individual streams from each user. 
Since the performance gains from \sysname increases with more users, 
we expect it to work well for 4 or more users. We do not test those cases for
brevity.
\begin{table}[t] 
	\caption{Evaluation in 2-Person Mixtures (SS)} 
	\label{tab:sep2}  
	\begin{threeparttable}[t]
	\centering
	\begin{tabular}{ p{2mm}| p{17mm} || c | c | c | c || c | c | c
			| c }
		\hlineB{4}
		& & \multicolumn{4}{c ||}{2-person mix (clean)}
		& \multicolumn{4}{c  }{2-person mix (noisy)} \\   
		\hline                       
		& Model & \metricone &
		\metrictwo & \metricthree & \metricfour & \metricone & \metrictwo &
		\metricthree & \metricfour \\
		\hline
		\multirow{5}{*}{\rotatebox[origin=c]{90}{\parbox[c]{5mm}{\centering
					Seen}}} 
		& Input             & 0.2 	      & -0.4  		& 0.71 		& 1.71 		& -1.7		& 0.3  		 & 0.61 	 & 1.37 	 \\    
		& ConvTasNet & 11.3 	      & 18.5 		& 0.87 		& 2.53		& 6.1       & 16.8  	 & 0.77 	 & 1.78 	 \\
		& SudoRMRF          & 10.9 	      & 15.4  		& 0.84		& 2.60 		& 4.7 		& 16.4  	 & 0.68 	 & 1.77 	 \\
		& DPRNN             & 13.5 	      & 21.5  		& 0.91 		& 2.63 		& 8.9  		& 20.3 		 & 0.81 	 & 1.96 	 \\
		& \sysname          & \bd{15.4 }  & \bd{23.6 }  & \bd{0.94} & \bd{2.83} & \bd{10.9} & \bd{23.3} & \bd{0.85} & \bd{2.10} \\
		\hline
		\multirow{5}{*}[-0.4ex]{\rotatebox[origin=c]{90}{\parbox[c]{10mm}{\centering
					Unseen}}} 
		& Input     & 0.0		& 0.53  	 & 0.70 	 & 1.62 	 & -1.8 	 & 0.30 	 & 0.60 	 & 1.39 	 \\    
		& ConvTasNet  & 9.5 		& 16.0   	 & 0.84 	 & 2.38 	 & 5.2  	 & 15.0 	 & 0.72 	 & 1.67 	 \\
		& SudoRMRF  & 6.2 		& 11.5   	 & 0.76 	 & 2.13 	 & 1.0  	 & 13.0 	 & 0.60 	 & 1.39 	 \\
		& DPRNN     & 10.8 		& 18.1   	 & 0.86 	 & 2.38 	 & 7.0  	 & 17.3 	 & 0.75 	 & 1.83 	 \\		
		& \sysname & \bd{14.5 }& \bd{22.3 } & \bd{0.92} & \bd{2.70} & \bd{10.3} & \bd{22.5 }& \bd{0.83} & \bd{2.05} \\
		\hlineB{4}
	\end{tabular}
	\end{threeparttable}
\end{table}

\begin{table}[t] 
	\caption{Evaluation in 3-Person Mixtures (SS)} 
	\label{tab:sep3}  
	\centering     
	\begin{tabular}{ p{2mm}| p{15mm} || c | c | c | c || c | c | c
			| c }
		\hlineB{4}
		& & \multicolumn{4}{c ||}{3-person mix (clean)}
		& \multicolumn{4}{c  }{3-person mix (noisy)} \\   
		\hline                       
		& Model & \metricone &
		\metrictwo & \metricthree & \metricfour & \metricone & \metrictwo &
		\metricthree & \metricfour \\
		\hline
		\multirow{3}{*}{\rotatebox[origin=c]{90}{\parbox[c]{5mm}{\centering Seen}}} 
		& Input     & -3.2 	      & -2.8  		& 0.60 		& 1.37		& -4.2 		& -2.8 		 & 0.55 	 & 1.30 	 \\    
		& DPRNN     & 7.2  	      & 14.0  		& 0.81 		& 1.95 		& 4.9    	& 15.7		 & 0.74 	 & 1.68 	 \\
		& \sysname & \bd{11.6}  & \bd{19.4 }  & \bd{0.88} & \bd{2.31} & \bd{9.3} & \bd{19.2} & \bd{0.83} & \bd{1.96} \\
		\hline
		\multirow{3}{*}{\rotatebox[origin=c]{90}{\parbox[c]{8mm}{\centering Unseen}}} 
		& Input     & -3.2      & -2.8       & 0.58 	 & 1.37 	 & -4.2 	 & -2.8   	 & 0.54 	 & 1.31      \\    
		& DPRNN     & 4.2  		& 10.2  	 & 0.73 	 & 1.72 	 & 2.6   	 & 12.5	     & 0.66 	 & 1.55 	 \\		
		& \sysname & \bd{10.7 }& \bd{18.2 } & \bd{0.86} & \bd{2.21} & \bd{8.6} & \bd{18.2 }& \bd{0.81} & \bd{1.90}  \\
		\hlineB{4}
	\end{tabular}
\end{table}

\subsection{Comparison with Audio Only Baselines}
As mentioned previously, introducing another modality has many benefits, such as
guiding the loss function at the beginning of training to solve permutation
problem and estimating the number of sources. To that end, in Fig.
\ref{fig:learningcurve}, we compare the loss values on training and validation
sets. As shown, the audioradio
system has a much steeper learning curve at the beginning, along with a better
convergence point.

\begin{figure}[t]
	\centering
	\includegraphics[width=0.7\columnwidth]{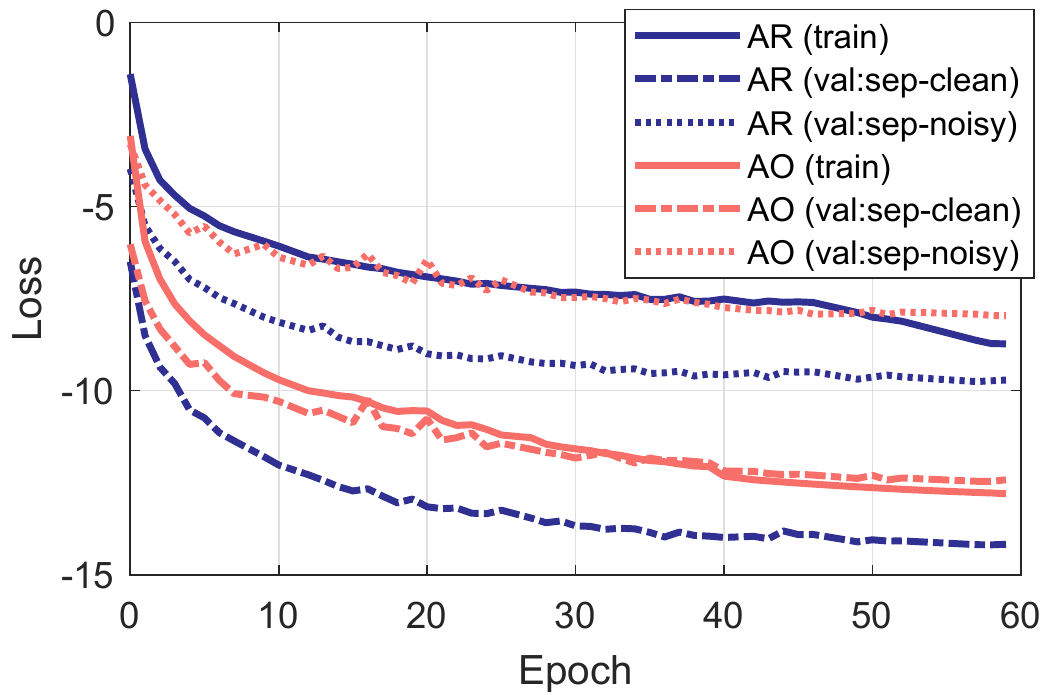}
	\caption{Learning curve for audio-only (AO) and audioradio (AR) for separating 2-person mixtures}
	\label{fig:learningcurve}
\end{figure}

\begin{figure}[t]
	\centering
    \subfloat[
		Output of sound separation
		\label{subfig:audioonlycomparison-abs}]
		{%
      \includegraphics[width=0.80\columnwidth]{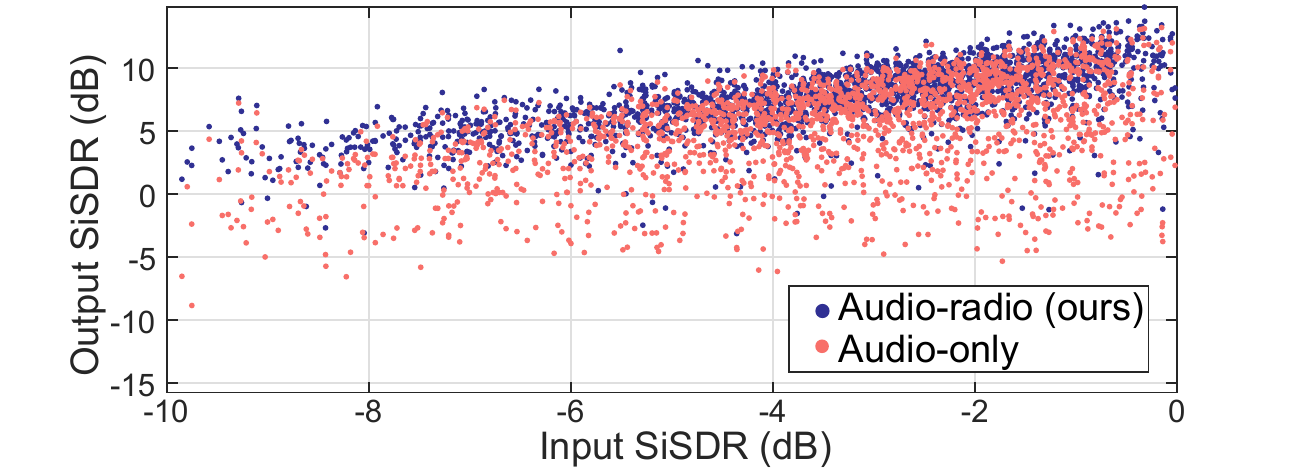}
    }
    \\
    \subfloat[Relative gains from radio channel
	\label{subfig:audioonlycomparison-relative}]
	{%
      \includegraphics[width=0.80\columnwidth]{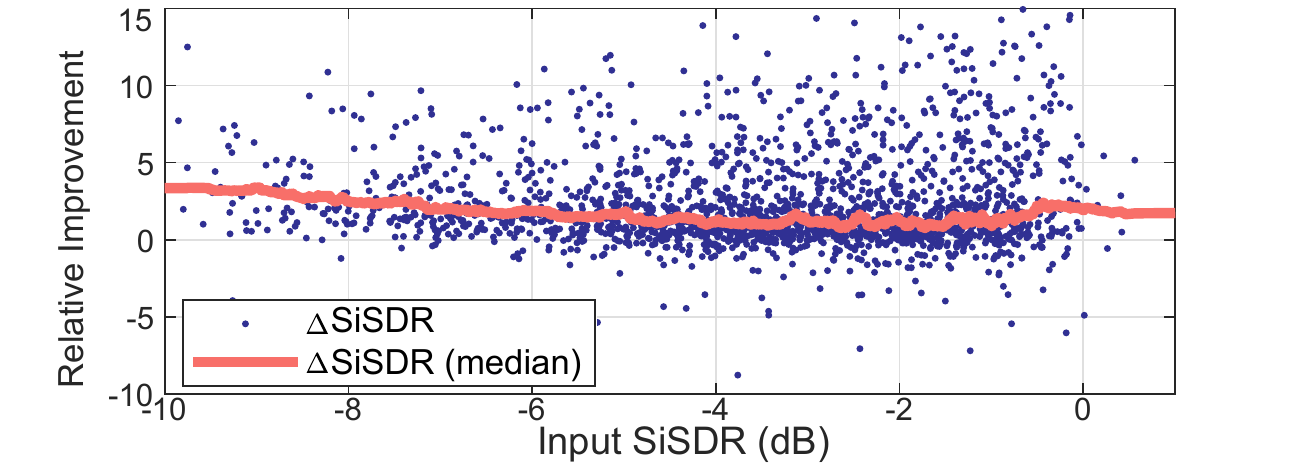}
	}
    \label{fig:audioonlycomparison}
    \caption{Comparison of \sysname with audio-only baseline in 2-person noisy mixture}
\end{figure}

\begin{figure*}[!ht]
	\subfloat[
	Distance setting
	\label{subfig:experiments-distance}]
	{%
		\includegraphics[width=\subfigwidthfour\textwidth]{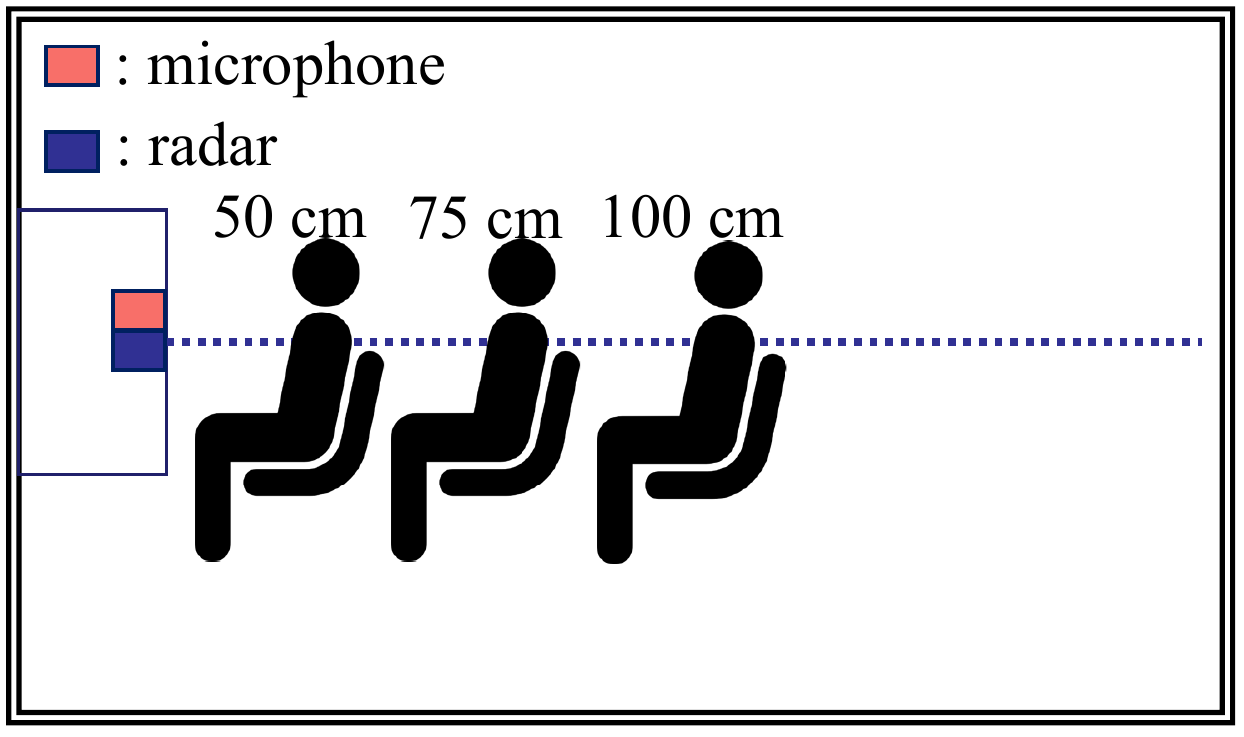}
	}
	\hfill
	\subfloat[Incident angle setting\label{subfig:experiments-angle}]{%
		\includegraphics[width=\subfigwidthfour\textwidth]{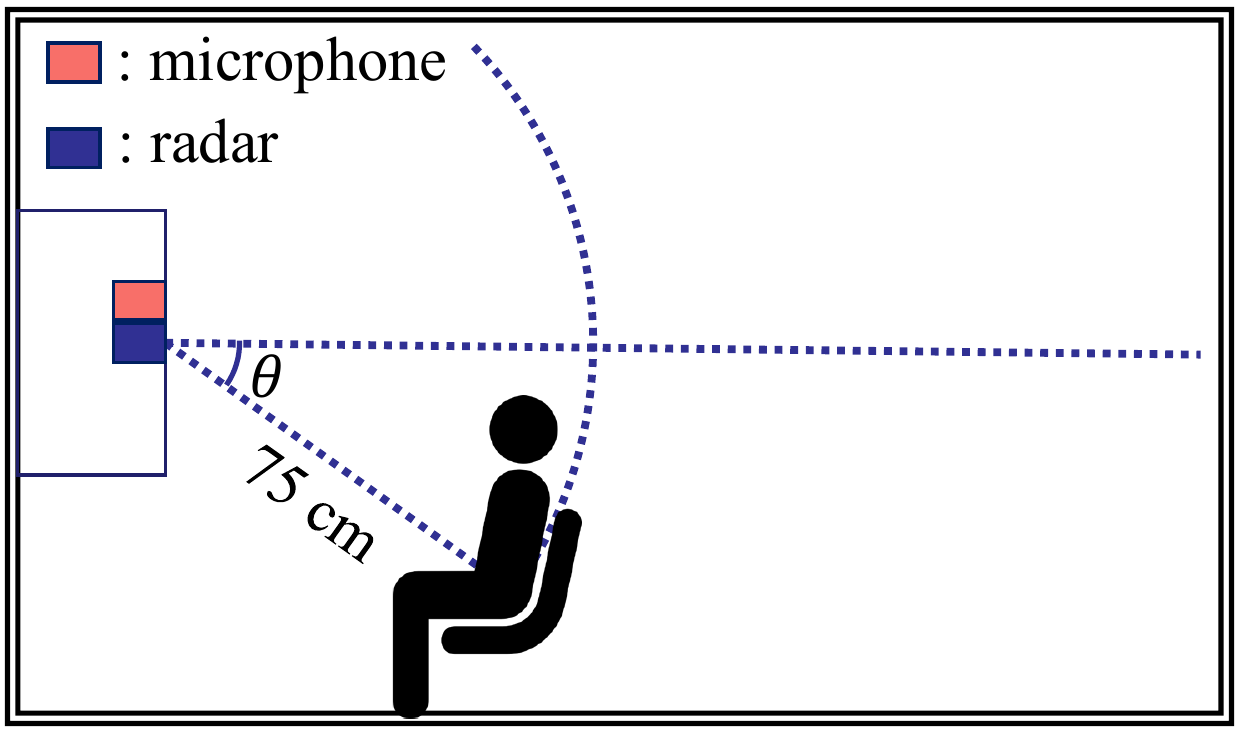}
	}
	\hfill
	\subfloat[Head orientation setting\label{subfig:experiments-orientation}]{%
		\includegraphics[width=\subfigwidthfour\textwidth]{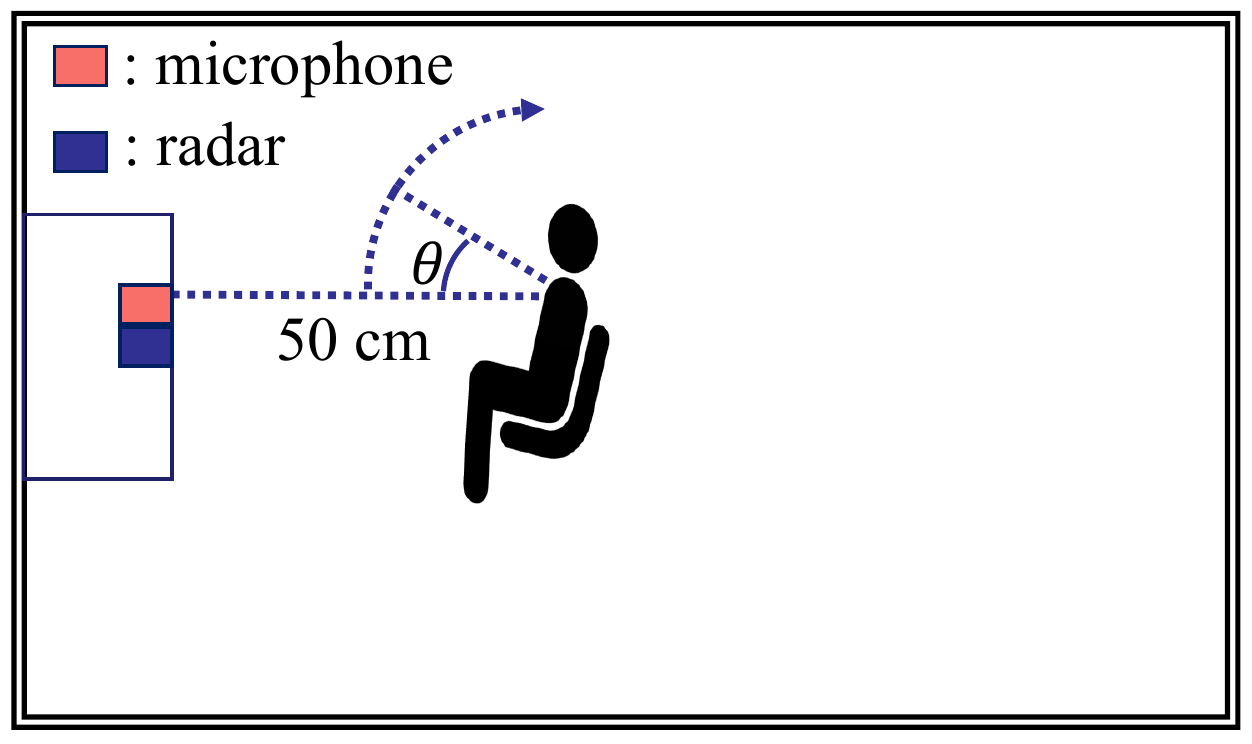}
	}
	\hfill
	\subfloat[Multi-user setting\label{subfig:experiments-otherroom}]{%
		\includegraphics[width=\subfigwidthfour\textwidth]{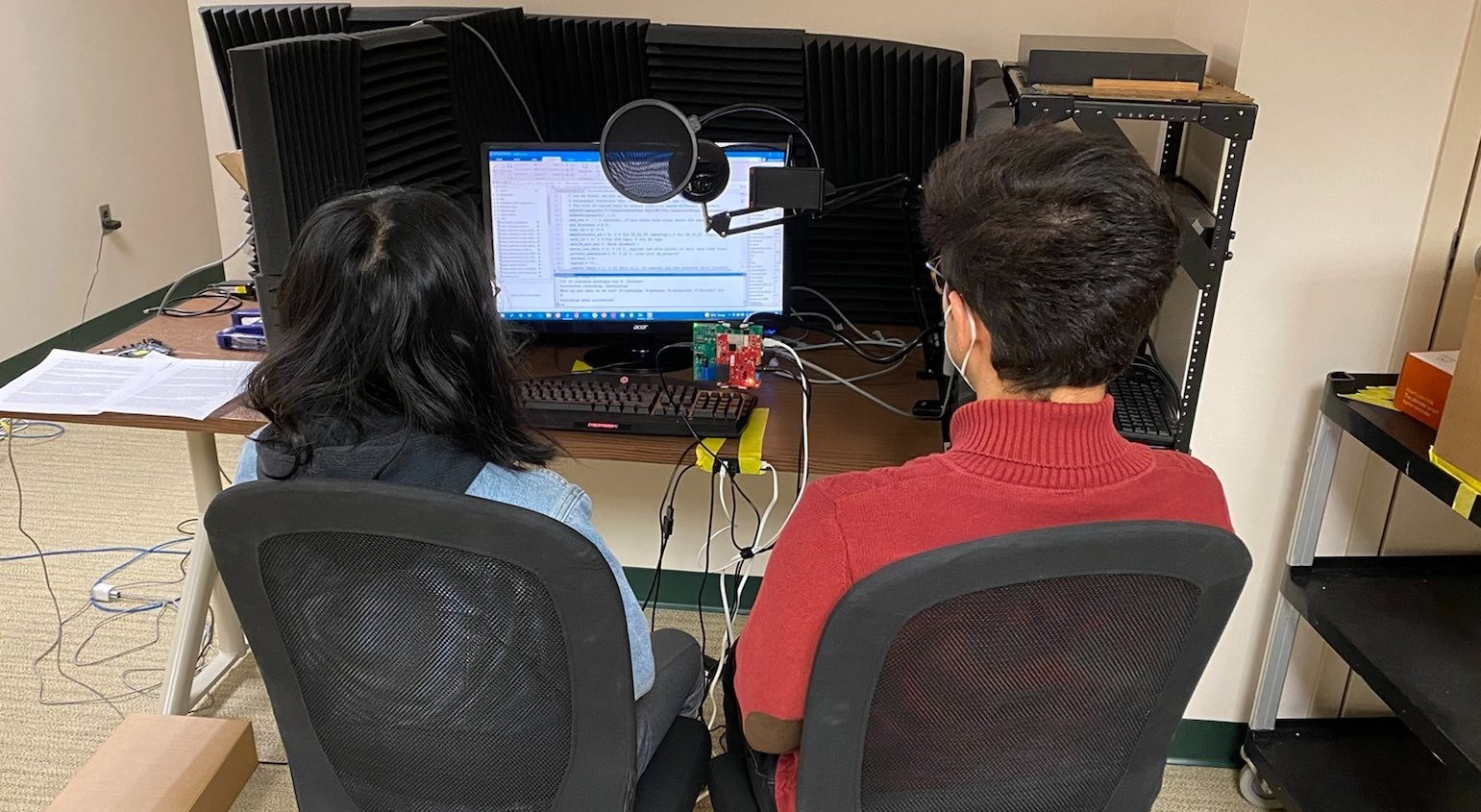}
	}
	\label{fig:experiments}
	\caption{Multiple experimental settings}
\end{figure*}

\begin{table*}[t]
	\caption{Performance with respect to multiple experiments of sources} 
	\label{tab:big-table} 
	\centering
	\begin{tabular}{l || c  c  | c  c  | c c || c c | c c | c c | c c  || c c | c c | c c  }
		\hlineB{4}
		Exp 
		& \multicolumn{6}{ c ||}{Distance}  
		& \multicolumn{8}{c ||}{Orientation}
		& \multicolumn{6}{c }{Head Orientation} \\
		\hline
		Case 
		& \multicolumn{2}{c |}{50 cm}
		& \multicolumn{2}{c |}{75 cm}
		& \multicolumn{2}{c ||}{100 cm} 
		& \multicolumn{2}{c |}{$0^\circ$}
		& \multicolumn{2}{c |}{$15^\circ$}
		& \multicolumn{2}{c |}{$30^\circ$} 
		& \multicolumn{2}{c ||}{$45^\circ$} 
		& \multicolumn{2}{c |}{$0^\circ$}
		& \multicolumn{2}{c |}{$15^\circ$}
		& \multicolumn{2}{c }{$30^\circ$}
		\\ 	
		\hline
		Metric  & AO   &  \ar  &  AO  &  \ar   & AO   &   \ar   &  AO   &  \ar  &  AO  &  \ar   & AO   &   \ar     & AO  & \ar  & AO   &  \ar  &  AO  &  \ar   & AO   &   \ar  \\
		\hline
		SiSDR  	& 6.3  & 10.9 & 3.8  & 8.6  &  2.3  & 4.3    & 3.8  & 8.6  &  3.6   & 7.8   & 4.4  & 8.3  & 4.2    & 8.2  & 6.3  & 10.9 &   5.6  & 9.8  & 5.4  & 9.3  \\
		SIR		& 12.5 & 18.3 & 9.9    & 15.6 & 8.7  & 9.8   & 9.9  & 15.6 &  9.6   & 14.8 & 10.6 & 15.1  & 10.2   & 15.6 & 12.5 & 18.3 &   11.7 & 16.8 & 11.5 & 16.3 \\
		STOI 	& 0.83 & 0.93 & 0.79   & 0.90 & 0.74 & 0.81  & 0.79 & 0.90 &  0.78  & 0.89 & 0.79 & 0.89  & 0.78   & 0.88 & 0.83 & 0.93 &   0.80 & 0.90 & 0.79 & 0.89 \\
		PESQ    & 2.17 & 2.61 & 1.97   & 2.42 & 1.79 & 2.00  & 1.97 & 2.42 &  1.91  & 2.32 & 2.00 & 2.33  & 2.02   & 2.37 & 2.16 & 2.61 &   2.11 & 2.46 & 2.10 & 2.43 \\
		\hlineB{4}
	\end{tabular}
\end{table*}

Furthermore, in Fig. \ref{subfig:audioonlycomparison-abs}, we compare the output
SiSDR of \sysname with its audio-only baseline. As shown, our proposed method
is superior to the audio-only baseline, and the performance gains are consistent
through different input SiSDR levels. 
To investigate the consistency of
audioradio system over audio, we plot the differential gain in terms of SiSDR in Fig.
\ref{subfig:audioonlycomparison-relative} from the radio channel. 
To characterize the incorrect associations, we check the amount of
samples with $\Delta (DB_i)<-3$ is $1.03\%$, indicating correct physical
association of sources for $98.97\%$ of the time. 

\vspace{-0.1in}
\subsection{Impact of Experiment Setting}
We further evaluate the performance of \sysname in varying settings, conducted
in a different location than the original data collection location. Since it is
difficult to \textit{simulate} the extracted radio signals from different
environmental scenarios, we collect data at
a variety of settings. For example, to test the effect of distance, 
we collect multiple
user data at different distances, (e.g. 75cm), 
and create mixtures from that
location. We normalize input
data streams to the same loudness levels for a fair comparison, 
although minor differences between each
setting is inevitable.
In order to show improvements, we present each settings'
performance along with the audio-only baseline, and show how \sysname preserves
a better performance in those settings. For presentation, we refer \sysname as
the audioradio (AR) method, 
whereas baseline DPRNNTasNet is noted as audio-only (AO) method.
As shown, \sysname mostly outperforms audio-only baseline with 4dB improvement
in our dataset, which includes unseen and same-speaker mixtures. 
This evaluation is done with clean mixtures for consistency, 
although we have observed similar gains in noisy
mixtures as well.

\subsubsection{Distance}
\label[]{subsec:distance}
First, we evaluate the effect of distance on the signal separation tasks, as
illustrated in Fig. \ref{subfig:experiments-distance}. 
As shown in Table \ref{tab:big-table},
\sysname can work
robustly until the speakers are 1m away from the device, and preserve the gains
compared to the audio-only baseline. The performance for both cases
decrease, which is due to training dataset being captured from a short distance 
only.
As the distance increases, the received audio signals change due to the room impulse
response and microphone nonlinearity, which is a phenomenon used for
coarse source distance estimation with microphones recently
(\eg~\cite{vesa2009binaural,yiwere2020sound}).
We note that, the performance gains from radio channel does not
decrease much from 0.5m and 1m, and the main bottleneck for lower performance is
the variety of audio data. A high-performance system can be built by capturing 
more diverse \textbf{audio} data. 

\subsubsection{Orientation}
\label[]{subsec:orientation}
Second, we ask the users to sit 0.75m away from the device and change their
orientation to explore the practical area of sensing, as illustrated in Fig.
\ref{subfig:experiments-angle}. 
We
realize that \sysname can work until $45^\circ$, without any performance
decrease, as presented in orientation columns of Table \ref{tab:big-table}. 
The gains from the
audioradio system are consistent (\eg~$\sim$4dB in SiSDR) through each
setting, showing the effectiveness in
modeling of the radio stream. Furthermore, this observation
is consistent with that of distance, as a different deviation angle from the
microphone does not create any distance-based nonlinearity, although it
reduces the radio-reflection SNR.

\subsubsection{Head Orientation}
\label[]{subsec:incidentangle}
Third, we ask users to sit at 0.5m, and rotate their heads from 0
degrees to 15 and 30 degrees, as shown in Fig. \ref{subfig:experiments-orientation}. For example, if a user sits in front of a laptop
or monitor, they would naturally swing their head to see different content on
the screen and 30 degrees of head rotation at 0.5m enables them to see the
entire
area of a big screen. Furthermore, if \sysname is using lip motion, instead of
vocal folds vibration, we would expect the results to deteriorate quickly.
The results are presented in the head orientation column of Table
\ref{tab:big-table}, which indicates that \sysname is robust
to changes in head orientation, even though the
training procedure does not include explicit head-rotation data. 

\subsubsection{Distortion}
\label[]{subsec:motion}
Fourth, we ask users to perform a variety of distortions. First, we
ask users to perform motions in front of the
radar while speaking. To have the experiments controlled, we ask the users to
move their heads up and down, left-to-right and back-and-forth naturally, as it can
happen during speech. Next, we collect data with users wearing a mask, which
plays a role as an occlusion. 
As shown in Fig.
\ref{fig:distortion-environment-radio}, \sysname is not affected by the head
motion. 
Furthermore, 
unlike certain visual
enhancement methods which lose their advantage with occlusions (as noted in
\cite{afouras2019my}),
\sysname is robust against wearing a mask and can preserve
the improvements compared to the audio-only method.
This
is due to the fact
that vocal
folds vibration are extracted from the body and throat, not from the face.
Similar improvements with respect to STOI (\eg from 0.8 to 0.9), and PESQ (\eg
from 2.1 to 2.5) are also observed, but not reported in the figures.

\begin{figure}
    \subfloat[
		SiSDR
		\label{subfig:distortion-sisdr}]
		{%
      \includegraphics[width=0.5\columnwidth]{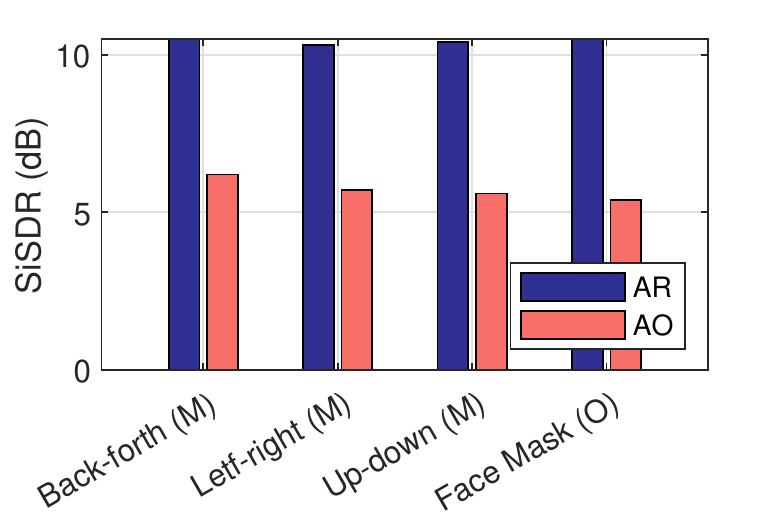}
    }
    \subfloat[SIR
	\label{subfig:distortion-sir}]
	{%
      \includegraphics[width=0.5\columnwidth]{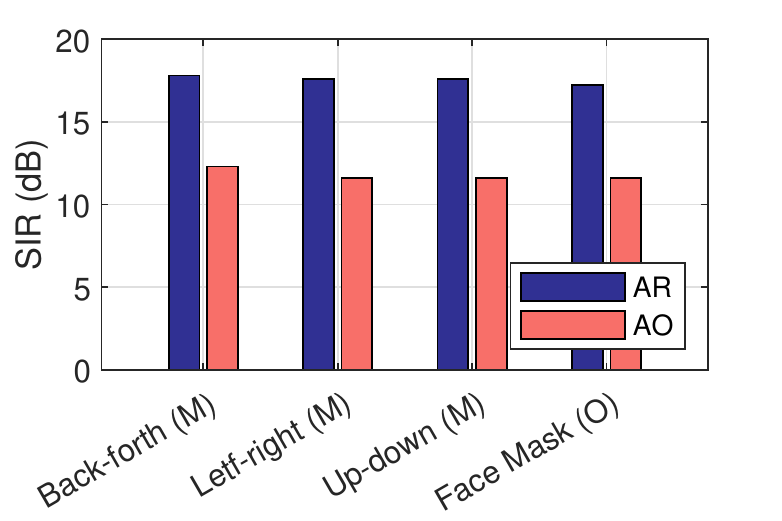}
	}
    \label{fig:distortion-environment-radio}
    \caption{Performance when there is motion (M) of the user, or occlusion (O).}
\end{figure}

\begin{table}[t]
	\caption{\textit{In the Wild} Experiment Results}  
	\label{tab:real-experiments-multiuser}  
	\centering     
	\begin{tabular}{l | c  c  c | c  c  c}
		\hlineB{4}
		Case 
		& \multicolumn{3}{c |}{Speech Enhancement}
		& \multicolumn{3}{c }{Speech Separation} \\    
		\hline
		Metric  & Clean & AR &  Noisy  &    Clean  &  AO  & AR \\
		\hline
		WER  	& 14  & 45 & 63 &  20    &  61    &  55   \\
		CER 	& 8   & 32 & 54 &  11    &  50    & 40   \\
		\hlineB{4}
	\end{tabular}
\end{table}

\subsection{Case Study in the Wild}
\label[]{subsec:realworld}
In this experiment, we ask multiple users to sit within the same room, and test
speech enhancement and separation in the wild, as shown in Fig.
\ref{subfig:experiments-otherroom} for the multiple speaker case. 
Although making a real-world system based on multimodal sensing, and
end-to-end deep learning frameworks involve additional challenges due to
Lombard effect \cite{michelsanti2019deep}, potential interference, and possible
covariate shift in the neural network layers, we try to explore whether there
would be improvements compared to an audio-only system.
We
ask a user to read Rainbow and Arthur passages (details in
\cite{xu2019waveear}), and play background noises from a pair of speakers.
Since this experiment does not have the ground truth clean
signals, we only evaluate the performance in terms of word-error-rate, and
character-error-rate. To have a fair comparison, 
we ask the users to read the same material in another quiet
environment and capture the performance in that setting. We use Google's
speech-to-text engine without any model adaptation to construct transcripts.
As our speakers are
not native speakers, and the \sysname is implemented with telephone-quality
speech (8 kHz), the overall error rate is higher. On the other hand, as presented in
Table \ref{tab:real-experiments-multiuser}, \sysname can enhance and separate
multi-person mixtures and outperform the audio-only baseline for speech
separation.
We also provide example files on our webpage at \url{https://zahidozt.github.io/RadioSES/}.

\vspace{-0.1in}
\subsection{Noisy and Partial Input Data}
\label[]{subsec:partialinput}
In this experiment, we corrupt input signals by adding noise and zero-padding,
which helps us to gain insight into the performance changes when people are
further 
away, or when there is package loss in the system.
These experiments are done with the first 3-seconds of the audio
streams, as longer audio streams already require some zero-padding or
overlapping block processing. 

\head{Noisy data:} We add white Gaussian noise to obtain 
radar data at
varying SNRs from 20 to -10  
dB levels, and report the performance metrics in Table
\ref{subfig:noisy-radio}. At larger distances,
radio signals are expected to be noisy, and this experiment explores when
the radio signals are still useful. \sysname outperforms
audio baseline, until a radio SNR of $-5 dB$. When the radio signal has further
noise, similar performance as the audio baseline is achieved. 
This experiment indicates that there is great potential for \sysname at larger distances.

\head{Partial input:} In this experiment, we zero pad the radio streams 
to reduce the available radar stream duration and test input radio durations of $2$s, $1.5$s, $1$s, and $0.5$s.
Such configurations can be
used when there are power requirements or 
package loss in the radio stream. As shown in
Figure \ref{subfig:shorter-radio}, \sysname can still help with speech separation
tasks and improve the performance, compared to the audio-only baseline, 
when there is at least 1s of signal (i.e., 33\%), in terms of
perceptual quality. 
\sysname system performs better than the audio-only
baseline with respect to all inputs after 1.5s of inputs. This indicates that
for power-constrained settings, \sysname can be operated with a duty-cycle less
than $50\%$, and can still bring performance improvements, along with the
aforementioned benefits of source association.

\begin{figure}
    \subfloat[
		Noisy radio inputs. $\infty$ represents undistorted
		radio signals, which actually includes device noise.
		\label{subfig:noisy-radio}]
		{%
      \includegraphics[width=0.5\columnwidth]{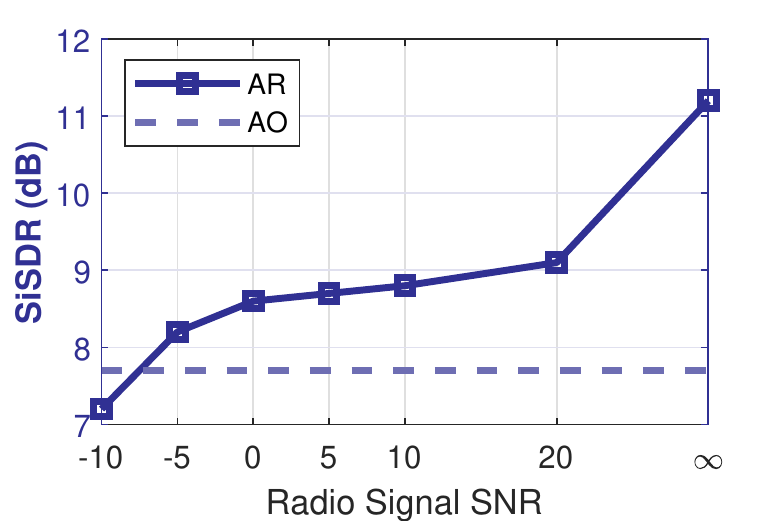}
    }
    \subfloat[Shorter radio inputs
	\label{subfig:shorter-radio}]
	{%
      \includegraphics[width=0.5\columnwidth]{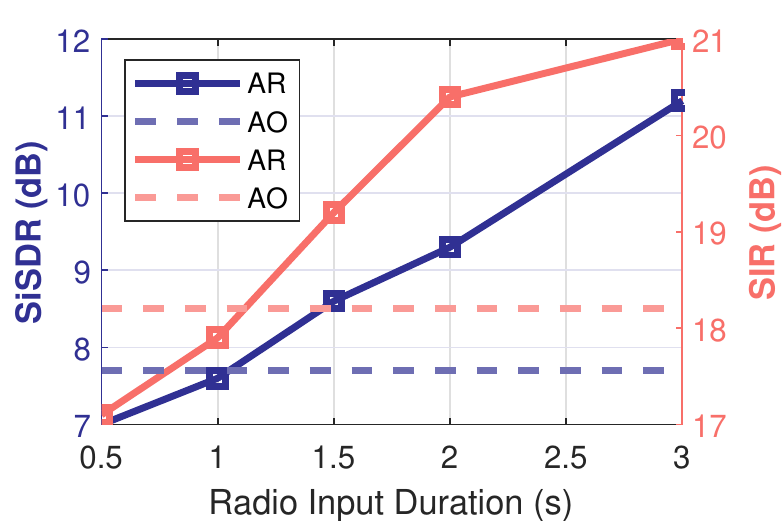}
	}
    \label{fig:distorted-radio}
    \caption{Performance for distorted radio inputs. Dashed lines represent the performance of the audio-only baseline}
\end{figure}

\subsection{Partial Detection}
\label[]{subsec:partialdetection}
Although having speakers outside the FoV of the radar is not our key focus in \sysname, 
we explore the limits of \sysname in such a mode of
operation, by allowing one speaker to be outside the FoV.
This setup requires use of alternative approaches to estimate the number of 
speakers, as
the radio-based methods will output fewer people 
(In practice, we may still use radio-based estimation by leveraging temporal 
information). 
We zero pad a radio stream to simulate no information from the outside user, and
investigate whether \sysname can benefit from having partial information. 
We evaluate a single person's missing case, but an extension to two
missing people is also possible, with permutation-based methods.
As shown in Table \ref{tab:partial-detection},
\sysname can still outperform the audio baseline with a large margin,
and improve the
performance, with missing people. We do not observe much performance decrease in
2-person
noisy mixtures, when one person is outside. 
For 3-person mixtures, there is more decline, but the gap
between audio-only system is larger, and benefits of having the two other
radio signals are clear.

\begin{table}[b]
	\caption{Performance for partial detection of sources}
	\label{tab:partial-detection}  
	\centering      
	\begin{tabular}{l | c  c  c | c  c  c}
		\hlineB{4}
		Case & \multicolumn{3}{c |}{2-person (noisy)}
		& \multicolumn{3}{c }{3-person (noisy)} \\    
		\hline
		Metric  & AO   &  AR(1) & AR(2) & AO & AR(2) & AR(3) \\
		\hline
		SiSDR  	& 7.7  & 10.1 & 11.2 & 4.9  & 8.3  & 9.3  \\
		SIR		& 18.2 & 20.7 & 21.0 & 13.0 & 17.7 & 19.2 \\
		STOI 	& 0.74 & 0.81 & 0.81 & 0.74 & 0.81 & 0.83  \\
		PESQ    & 1.95 & 2.19 & 2.20 & 1.68 & 1.89 & 1.96  \\
		\hlineB{4}
	\end{tabular}
\end{table}

\subsection{Ablation Studies}
\label[]{subsec:ablation}
In this experiment, we train \dlmodule without several blocks to understand the
effect of each component. We use clean 2-person mixtures for our ablation study.
As shown in Table \ref{tab:ablation-studies}, we remove i) Radio DPRNN blocks
ii) Audio DPRNN blocks and iii) High-pass (HP) filter from the mask estimation.
In the last case, the audio stream is still used to encode the signal, in order not
to change the main structure of \sysname, but is not passed through any DPRNN
blocks. 

\section{Discussion}
\label[]{sec:discussion}

In this work, we propose \sysname to improve the robustness and performance of SES
tasks using radio modality. 
Despite promising results with \sysname, there are certain limitations and many
interesting directions to pursue further. 

\head{Other side channels:} Although in this work we assume the vibration
sources in the field-of-view of radio device to be from vocal folds only, radios
can also measure vibration
of other sources, such as guitars \cite{ozturk2021radiomic}, or 
machinery \cite{jiang2020mmvib}. These vibration sources usually
create some sound signature, and they can be used to estimate the sound from
each 
source separately, as done using cameras in \cite{zhao2018sound}.

\begin{table}[t]
	\begin{minipage}[t]{0.25\textwidth}
		\vspace{0pt}
		\caption{
			Ablation Study: {\rm Radio modality and HP Filter are essential parts of
			\dlmodule, whereas additional radio DPRNN blocks bring extra performance
			improvements.}
		 } \label{tab:ablation-studies}
	\end{minipage}\hfill
	\begin{minipage}[t]{0.22\textwidth}
		\vspace{0pt}	  
	  \begin{tabular}{ l   c }
		\hlineB{4}                  
		Model & SiSDR \\
		\hline
		\dlmodule   & 15.4 \\
		\myalign{r}{w/o Radio DPRNN} & 15.2 \\
		\myalign{r}{w/o Any Radio} & 13.5   \\
		\myalign{r}{w/o Audio DPRNN} & 4.8  \\
		\myalign{r}{w/o HP filter} & 0.1    \\
		\hlineB{4}
	\end{tabular}
	\end{minipage}
  \end{table}

\head{Microphone arrays:} \sysname uses a single microphone along with an mmWave
sensing device. On the other hand, it is also possible for \sysname to work with
a microphone array, and radio modality can still bring further improvements to
overall performance. Although beamforming in microphone arrays may indicate that
radio modality is unnecessary, it can fail in noisy or reverberant 
\cite{michelsanti2021overview} environments. Since
\sysname senses the vibration of the \textit{source}, it
can estimate the direction of the sound for robust beam-steering or can extract
the source vibration without any
reverberation for further improvement. Some recent work addresses this problem
in 
audiovisual domain \cite{tan2020audio}, and we believe similar contributions
using \sysname can be achieved in the future.

\head{Moving Speakers:} Currently, 
\sysname is designed to track bodies with the
assumption that they do not move significantly. This is usually a
common constraint in the relevant vital signs monitoring literature (breathing,
heart rate), although some recent work started addressing motion for
breathing \cite{zheng2021morefi}. A more thorough system should support
medium and high levels
of 
source motion. To that end, coherent combining of multiple vocal fold bins from person
point clouds (\eg~\cite{wang2021vimo}), or deep learning \cite{zheng2021morefi} can be some
interesting ideas to support 
multiple moving targets.

\head{Sensing Distance:} Our experiments indicate that \sysname can work
robustly until the speakers are 1m away from the device, and preserve the gains
compared to the audio-only baseline. The performance for both cases
decreases, which is due to the training \textit{audio} dataset being captured from a short distance.
However, the performance improvements from \sysname do not
decrease much with the distance. During our experiments,
we realized that raw signal SNR is still high at large distances (\eg~2.5m) for
people with
low pitch (\eg~males). To support all users,
we limited the practical range to 1m, 
much larger than the range of using ultrasound \cite{sun2021ultrase}.
Although not much radar signature can be captured from these bodies when they
are further away, they can still be
robustly detected, (\eg~as in vital sign monitoring), and
even the reduced number of high quality radio streams can still help to improve
the performance, as illustrated in \S\ref{subsec:partialdetection}.
Moreover, a 
different hardware can capture vocal folds vibration from $7m$
in \cite{chen2017detection}, or at $50m$ \cite{xiang2021acquisition}. We
believe \sysname can benefit from better hardware significantly, and a more
practical system can be built.

\head{Multipath Effects:} In our experiments, we consider cases with multiple
sources in front of the radar, and training data assumes perfectly clean 
radio streams for each person. However, in challenging conditions,
wireless sensing-based systems can have a strong multipath effect. Although
in mmWave bands, the effect is not as detrimental as 2.4/5 GHz, it can still
reduce the performance. We did not encounter this issue in our short-range
experiments, but it can be a limiting factor for long-range indoor sensing.
We plan to address this issue in the future by potentially
simulating multipath data. 

\head{Power Consumption and Cost:} Although our evaluation board costs $\$300$,
a single mmWave device can be purchased for $\$15$ from TI. Transmission power
of the device is 12 dBm ($\approx 16mW$) and the selection of radar parameters
result in a duty cycle of $7.3\%$, (\ie $\approx 1.2mW$). For comparison, the size of
these devices 
can go as small as $6mm\times6mm$ to fit in a phone \cite{soliphone}, 
and the power consumption of the radar in that phone is 
$1mW$
\cite{soliphone}. Furthermore, \sysname does not require capturing the entire
signal duration (\S\ref{subsec:partialinput}) and based on the application,
lower power consumption can be achieved by reducing the duty cycle further down. As there are already devices with
continuous mmWave sensing capabilities, we believe \sysname is
feasible to be integrated with smart
devices, 
and this
work introduces a new application. 

\vspace{-0.1in}

\section{Conclusion}
\label[]{sec:conclusion}
We present \sysname, a joint audioradio speech 
enhancement and separation system using mmWave sensing. 
It improves the performance of
existing audio-only methods with the help of radio modality and achieves similar
improvements as audiovisual systems, with further
benefits in computation complexity and privacy. 
Furthermore, \sysname 
can detect the number of sources in the environment, and associate outputs with
the physical speaker locations,
 all being challenging problems in audio-only
domain. 
Real-world experiments show that \sysname outperforms the state-of-the-art
methods considerably (\eg~3 dB SiSDR improvements in 2-speaker mixtures w.r.t.
audio-only baseline), 
demonstrating the great potential of audioradio SES.

\bibliographystyle{ieeetran}
\bibliography{mybibliography_radarsound}

\end{document}